\documentclass[journal]{IEEEtran}

\usepackage{graphicx}
\usepackage{wrapfig}
\usepackage{graphicx}
\usepackage{subfigure}
\usepackage{float}
\usepackage{adjustbox}
\usepackage{tabu}
\usepackage[justification=centering]{caption}
\usepackage{tabularx}
\usepackage{multirow}
\usepackage[belowskip=-5pt,aboveskip=0pt]{caption}
\usepackage{array}
\newcolumntype{P}[1]{>{\centering\arraybackslash}p{#1}}
\newcolumntype{M}[1]{>{\centering\arraybackslash}m{#1}}
\newcommand{\PreserveBackslash}[1]{\let\temp=\\#1\let\\=\temp}
\newcolumntype{R}[1]{>{\PreserveBackslash\raggedleft}p{#1}}
\usepackage{tabulary,booktabs}
\usepackage{lipsum}
\usepackage{blindtext}
\usepackage{amsmath,amssymb}

\usepackage{hyperref}

\usepackage{enumitem}

\usepackage[linesnumbered, algo2e, ruled,norelsize]{algorithm2e}
\usepackage[table]{xcolor}
\usepackage{cite}

\SetCommentSty{mycommfont}

\usepackage{gensymb}
\DeclareMathOperator{\E}{\mathbb{E}}

\usepackage{authblk}

\title{Lossless Point Cloud Geometry and Attribute Compression Using a Learned Conditional Probability Model}
\author{Dat Thanh Nguyen, Andr\'e Kaup \vspace*{-.35cm} 
\thanks{This work was funded by the Deutsche Forschungsgemeinschaft (DFG, German Research Foundation) under Grant SFB 1483 – Project-ID 442419336. Corresponding Authors: D.T.Nguyen (\href{mailto:dat.thanh.nguyen@fau.de}{dat.thanh.nguyen@fau.de}); A. Kaup (\href{mailto:andre.kaup@fau.de}{andre.kaup@fau.de}).}}
\affil{Chair of Multimedia Communications and Signal Processing  \\ Friedrich-Alexander-Universität Erlangen-Nürnberg (FAU) \\ Erlangen, Germany}
\begin{document}

\makeatletter
\def\ps@IEEEtitlepagestyle{
  \def\@oddfoot{\mycopyrightnotice}
  \def\@evenfoot{}
}
\def\mycopyrightnotice{
  {\footnotesize
  \begin{minipage}{\textwidth}
  \centering
  Copyright~\copyright~2023 IEEE.  Personal use of this material is permitted. However, permission to use this material for any other purposes must be obtained from the IEEE by sending an email to pubs-permissions@ieee.org.
  \end{minipage}
  }
}
\maketitle
\begin{abstract}
In recent years, we have witnessed the presence of point cloud data in many aspects of our life, from immersive media, autonomous driving to healthcare, although at the cost of a tremendous amount of data. In this paper, we present an efficient lossless point cloud compression method that uses sparse tensor-based deep neural networks to learn point cloud geometry and color probability distributions. Our method represents a point cloud with both occupancy feature and three attribute features at different bit depths in a unified sparse representation. This allows us to efficiently exploit feature-wise and point-wise dependencies within point clouds using a sparse tensor-based neural network and thus build an accurate auto-regressive context model for an arithmetic coder. To the best of our knowledge, this is the first learning-based lossless point cloud geometry and attribute compression approach. Compared with the-state-of-the-art lossless point cloud compression method from Moving Picture Experts Group (MPEG), our method achieves 22.6\% reduction in total bitrate on a diverse set of test point clouds while having 49.0\% and 18.3\% rate reduction on geometry and color attribute component, respectively. 
\end{abstract}


\section{Introduction}
%
%
%
%

\label{sec:intro}
\IEEEPARstart{T}{he} advances in 3D capturing technology enable highly detailed scenes with six degrees of freedom being stored in the form of a point cloud. As a result, point cloud data has been becoming one of the most important data structures in many 3D visual applications such as multimedia, autonomous driving, healthcare, etc. However, each point cloud contains millions of 3D points in $x, y, z$ representation associated with attribute information. This produces extremely large files which are difficult to store, retrieve and transmit. In addition, because of the irregular sampling and sparsity of this data structure, point cloud coding is even more challenging compared to image/video coding. Therefore, an effective point cloud compression (PCC) method is highly desirable. 
\par One of the well-known standardization groups for point cloud coding is the Moving Picture Expert Group (MPEG), which proposed two point cloud compression methods \cite{graziosi2020overview,8571288,jang2019video,8945224}: Video-based PCC (V-PCC) and Geometry-based PCC (G-PCC). V-PCC projects 3D point cloud data to 2D planes and utilizes conventional 2D video coding standards to compress the generated videos. On the other hand, G-PCC processes and compresses point clouds using an octree representation with independent coders for geometry and attribute. Color attributes are  encoded using color transforms such as Region Adaptive Hierarchical Transform (RAHT) \cite{graziosi2020overview}, Predicting Transform or Lifting Transform \cite{de2016compression}. In this paper, we also propose an efficient scheme to compress both 3D point cloud geometry and attribute. 

\par Specifically, we focus on static voxelized point clouds which can be represented in either octree, voxel, or point representation. Point or sparse representation is the raw representation produced by most 3D capturing devices and the scenes can be reconstructed directly from this representation. A voxel block is obtained by dividing the 3D space into a fixed number of cubes per dimension which defines the resolution of a point cloud (512 cubes = 9 bit-depth). Each unit cube is then called a voxel and a voxel is occupied if it contains at least one point. Besides the occupancy state, a voxel can contain attribute information such as color, reflectance, velocity, etc. Usually, very few voxels are occupied and carry attribute information, most of the voxels are empty which creates a significant amount of redundancy. By recursively dividing the voxel block into eight sub-blocks and using a binary bit to signal the occupancy state of each sub-block, we obtain an octet representing the occupancy of each node. The octree is fully constructed when the unit sub-block is reached. Octree decomposition efficiently removes the empty space within PCs by stopping splitting on empty nodes and thus providing smaller raw bitrate compared to a voxel representation. However, an octree is not a 3D-based representation and thus the geometric information (e.g. planes, curves, lines, etc.) is lost.
\par Similar to image and video, the spatial and temporal redundancies within point clouds can be exploited for efficient point cloud coding methods. However, it is not a straightforward extension to bring the powerful and popular signal processing tools such as discrete cosine transform (DCT), Fourier transforms, and convolution to 3D space because of the sparsity, irregularity, and high dimensionality of point cloud data. 

\par We address the mentioned representation issues by bringing sparse tensor representation and sparse convolution \cite{choy20194d} to point cloud coding tasks. First, sparse convolution has been specifically designed for high-dimensional signals which can process them efficiently with the least computational cost. Second, unlike octree representation, in sparse tensor representation, the geometric information of a point cloud (e.g. planes, curves, lines, etc.) and attribute correlation are preserved and can be exploited directly by sparse convolution. Note that our focus is to compress 3D point cloud attributes (e.g., colors and velocities, etc.), and the color attribute is used as a representative case in our work, without generality loss.
\par We propose for the first time a learning conditional probability model-based method for lossless compression of the static point cloud which we refer to as CNeT. We estimate the probability distributions of each feature sequentially such that the latter feature is modeled conditionally on the previous features. Within each feature, we apply a similar sequential modeling method to estimate the probability distribution of each point. The conditional distribution is then used to model the context of a context-based arithmetic coder.
\par We investigate the benefits of the transform from RGB to luminance-chrominance color space in our point cloud compression method. We conduct a lossless transformation and encode the transformed color features in an order such that our context model can efficiently exploit the correlation between features and thus minimize the final bitrate. Besides, we also study the effect of Softmax-based and Mixture of Logistic-based distribution modeling in our point cloud compression task. Extensive experimental results show the superiority of our method over the state-of-the-art MPEG G-PCC lossless codec in terms of bits per point over a diverse set of point clouds from four different datasets. 
\par The rest of this paper is organized as follows. Section \ref{sec:stateoftheart} reviews existing coding schemes for 3D point clouds. The proposed compression method is discussed in Section \ref{proposedmethod}. Experiments, as well as comparisons and analysis are conducted in Section \ref{experiment}. Finally, the conclusion and future work is described in Section \ref{conclusion}.



\section{Related work}
\label{sec:stateoftheart}

\begin{figure*}
\centering
\captionsetup{justification=justified, singlelinecheck=false}
\includegraphics[width=.95\linewidth]{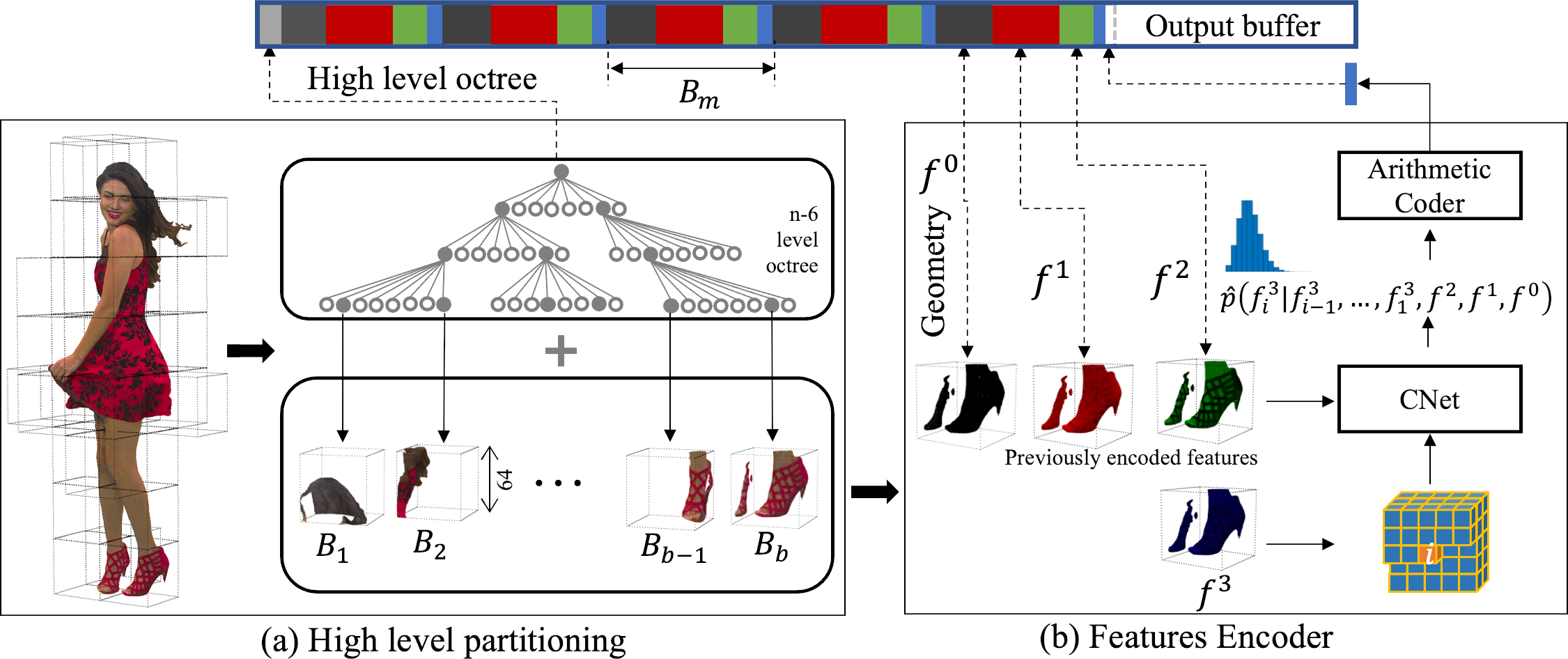}
\vspace{0.1cm}
\caption{Architecture overview of the proposed method. (a): a $n$ bit depth point cloud is partitioned down to the $n-6$ octree level, yielding $b$ occupied blocks of size $64\times 64 \times 64$ ($B_1$ to $B_b$). (b): Each block contains four features including geometry $f^0$ and three color features $f^{1-3}$. \textcolor{black}{We encode feature by feature sequentially from $f^0$ to $f^3$, the latter feature is encoded conditionally based on previously encoded features.} At this figure, the last feature $f^3$ is being encoded. The top bar shows the output bitstream, the high level octree is stored first, the feature bitstreams of each block $B_m$ are marked with corresponding colors. }
\label{fig:system overview}
\end{figure*}

\par Each point cloud contains geometry and attribute information in which geometry is usually encoded first, prior to color coding. As a result, we have seen two main parts in 3D point cloud compression: geometry and attribute compression.
\subsection{Point Cloud Geometry Compression}
\par Many of the existing point cloud geometry  compression  methods are based on trisoup representation or octree representation \cite{schnabel2006octree,graziosi2020overview, garcia2017context, garcia2018intra,garcia2019geometry,biswas2020muscle, huang2020octsqueeze,dricot2019adaptive}.  Trisoup-based coders represent geometry as a pruned octree plus a surface model which is approximated using 1 to 10 triangle meshes. Such surface modeling is included in the MPEG G-PCC coder, however, this is preferred for lossy dense point cloud coding. In contrast, an octree is a well-known representation for point cloud geometry compression because of its benefits in representing irregular point space. Octrees can be used for both lossy and lossless scenarios. The common approach for octree-based lossless geometry compression is utilizing the octree structure to provide better context for entropy coding. For example, the octree-based lossless coding method of the G-PCC coder relies on conventional methods to exploit local geometry information such as neighbour-dependent entropy context \cite{neighbor}, intra prediction \cite{intracodinggpcc}, planar/angular coding mode \cite{planarcodingmode,angularcodingmode}, etc. More context modeling-based approaches using the information from parent nodes, current node position, or previous time frame are proposed in \cite{garcia2017context, garcia2018intra, garcia2019geometry}. A learning-based context modeling method for lossless point cloud geometry compression on the octree domain has also been studied in \cite{huang2020octsqueeze, biswas2020muscle, kaya2021neural, fu2022octattention}. For instance, the authors in \cite{biswas2020muscle} model the probabilities of octree symbols by considering both higher-level geometry, previous frame geometric and intensity information using a deep neural network. \textcolor{black}{Similarly, the authors in \cite{kaya2021neural} proposed a learning-based geometry compression method - NNOC which leverage the 3D context from surrounding nodes to encode the octree. An attention mechanism-based method with sibling and ancestor context modeling (OctAttention) was proposed in \cite{fu2022octattention} to losslessly encode octree symbol sequences. }  

\par Recently, with the advances in deep learning techniques, voxel-based and sparse tensor-based approaches for both lossy and lossless point cloud geometry compression have emerged. In particular, the auto-encoder framework has been widely used for lossy coding \cite{theis2017lossy,quach2019learning,yan2019deep,wang2021multiscale,wang2021sparse,9287077}. The point cloud geometry is first projected onto lower-dimensional latent spaces, the latent spaces are then encoded by entropy coders with auto-regressive and/or hyperprior-based context modeling. In terms of lossless point cloud geometry coding, efficient auto-regressive voxel-based compression methods \textcolor{black}{(VoxelDNN)} have been proposed in \cite{nguyen2021learning,nguyen2021lossless} which was further developed to a multi-scale approach by the authors \cite{nguyen2021multiscale}. However, those voxel-based point cloud geometry compression methods require high computational complexity. Sparse tensor-based lossless geometry compression has been shown as a potential avenue with high efficiency and low computational cost \cite{nguyen2022learning,wang2021sparse}. \textcolor{black}{For instance, the authors in \cite{wang2021sparse} proposed an efficient PCC method (dubbed as SparsePCGC) which progressively downscales the geometry tensor and uses a SparseCNN-based Occupancy Probability Approximation module (SOPA) to predict occupancy probabilities over either single or multi prediction step.}
\subsection{Point Cloud Attribute Compression}
\par In addition to the well-developed geometry compression, many approaches have been devoted to point cloud color compression. One popular strategy is compressing attributes based on geometry-dependent transforms or transform-based methods \cite{7025414,7482691,8462684,4378368,gu20193d}. For instance, the proposed method in \cite{7025414} constructs a graph from the geometry and conducts a Graph Fourier Transform (GFT) over the attribute signal on the graphs before quantization and entropy coding the coefficients. However, GFT is inefficient due to the high computational cost introduced by eigenvalue decompositions. To alleviate this, the region-adaptive hierarchical transform (RAHT) was introduced in \cite{7482691} to compress point cloud attributes and achieved a better trade-off between performance and complexity. Specifically, a hierarchical subband transform that resembles an adaptive variation of the Haar wavelet is used to transform point cloud color attributes. The adaptive run-length Golomb-Rice (RLGR) arithmetic coder for the quantized transform coefficients is also introduced. By adding RAHT prediction, better entropy coding, and other improvements, RAHT became one of the core modules in the MPEG G-PCC reference software TMC13 \cite{TMC13}. In addition to RAHT, the lifting transform \cite{lifting} and the predicting transform \cite{predicting} are the transform methods for attribute compression in G-PCC. However, the predicting transform, which is an interpolation-based hierarchical nearest-neighbor prediction scheme, is typically used for lossless coding. \textcolor{black}{An interesting approach using geometry-guided sparse representation and graph-based method for lossy point cloud attribute compression has been proposed recently \cite{gu20193d, gu20203d, liu2021hybrid}. However, the encoding/decoding time of those methods are still impractical due to the generating/solving sparse representation step.} 

\par Point clouds can also be projected to 2D planes and then utilize 2D image/video compression standards to encode the generated images/videos \cite{7434610,9169853}. In \cite{7434610}, attributes are organized into $8\times8$ blocks using snake scanning before being projected to 2D images and compressed using JPEG. Similarly, MPEG V-PCC converts the point cloud data into a set of video sequences, the video sequences are then encoded using state-of-the-art video coding (e.g. HEVC) \cite{graziosi2020overview}. 

\par Neural networks have been adopted for geometry compression with some success but are still in the early phase for attribute compression. Recently, there has been an increase in learning-based point cloud attribute compression methods \cite{quach2020folding,alexiou2020towards,9447226,isik2021lvac,wang2022sparse}. In \cite{9447226}, the attribute is compressed by representing them as samples of a vector-valued volumetric function which is modeled by neural networks. Authors in \cite{quach2020folding} introduced a deep learning method for mapping 3D attributes to 2D images and take advantage of a 2D codec to encode the generated image. This mapping method, however, introduces distortion and high complexity. Another approach for lossy compression is adopting an auto-encoder framework that directly encodes and decodes point cloud attributes with the help of geometry on voxel grids \cite{alexiou2020towards}. An auto-encoder lossy compression method has also been proposed in \cite{wang2022sparse}. The authors utilize a sparse convolution and adaptive entropy model with auto-regressive and hyper prior to achieve better performance. \textcolor{black}{However, these approaches only achieve a similar performance to RAHT while RAHT is a sub-module in G-PCC and had been significantly improved with additional tools  (e.g. prediction, entropy coding, ...). } 
\par Most of the lossless attribute compression schemes are based on conventional methods \cite{huang2021hierarchical,yin2021lossless,song2021layer} including G-PCC attribute lossless compression \cite{graziosi2020overview}. For example, a hierachical bit-wise differential coding-based method (HBDC) for static point cloud attribute compression was recently proposed in \cite{huang2021hierarchical}. The authors conduct a breath-first traversal of the octree and encodes the residual between parent and child nodes when necessary. The HBDC method shows a maginal gain for attribute compression over G-PCC version 11 while being inefficient at geometry compression. A normal-based intra-prediction scheme for LiDAR point cloud attributes is proposed in \cite{yin2021lossless}. The authors improve the Prediction Transform of G-PCC \cite{predicting} to further exploit local similarity by introducing the normals of point clouds. However, the method only achieves minor improvements compared to the original lossless compression method of G-PCC. In this paper, we build an efficient adaptive context model for entropy coders using deep neural networks. To the best of our knowledge, this is the first learning-based lossless scheme for point cloud attribute compression. 

\section{Proposed Method}
\label{proposedmethod}

\subsection{Overview}
\label{overview}
\par To overcome the challenges of point cloud data including sparsity, irregularity, and high dimensionality, we adopt sparse tensor and sparse convolution \cite{choy20194d} into the PCC task. On the one hand, a sparse tensor allows us to store only the occupied 3D space using its coordinates and the associated features or attributes. On the other hand, sparse convolution, which was designed for sparse tensors, closely resembles conventional convolution and efficiently saves memory and computation cost. In our method, a point cloud can be represented as a sparse tensor $\{ C, F\}$ carrying a set of 3D coordinates $C=\{(x_i,y_i,z_i)\}_i$ and a set of associated features $F$. The feature set can include one occupancy state and Red, Green, and Blue features or other color features such as YCgCo or YUV with different bit depths.  For the generality, we use the notation $F=\{f^0,f^1,f^2,f^3\}$ for a feature set containing one occupancy channel $f^0$ and three color channels $f^{1-3}$. In this paper, we aim to losslessly encode all the features in $F$.
\par The most well-known and efficient method for lossless compression is entropy coding. A stream of symbols, which are drawn from a given set of symbols $X$, is encoded to a bitstream using the probability density function $p$. Encoders will be optimized to minimize the average bits per symbol $L$ or the bit-rate as:
\begin{equation*}
L=\sum_{x \in X} p(x)l(x)
\end{equation*}
where $l(x)$ is the code length for each symbol. Generally, the entropy coder assigns each symbol with a code word or a number interval which is corresponding to its probability. A higher probability produces a shorter codeword or large interval and thus minimizes $L$. According to Shannon's source coding theorem \cite{shannon1948mathematical}, the lower bound of $L$ is the entropy of the source: 
\begin{equation*}
\begin{split}
L \geq H(p)
&= \E_{x \sim X}\left[-\log(p(x) \right]\\
&=\sum_{x \in X} - p(x) \log p(x) 
\end{split}
\end{equation*}
However, in practice, the real data probability density function $p$ is unknown and is usually estimated by a probability model $\hat{p}$. Hence, the cross-entropy (CE) between $\hat{p}$ and $p$, which is the sum of $H(p)$ and a positive term - the Kullback–Leibler divergence $D_{KL}(p||\hat{p})$, is used as a sub-optimal lower bound for expected bits per symbol: 
\begin{equation} \label{eq:minimumbitrateeq}
\begin{split}
   \tilde{L} \geq H(\hat{p},p) 
   &= \E_{x\sim X}\left[-\log(\hat{p}(x) \right]\\
&=\sum_{x \in X} - p(x)\log \hat{p}(x) 
\end{split}
\end{equation}
\par In this paper, we minimize the lossless bitrate needed to encode the features by constructing a probability density model $\hat{p}$ that minimizes Eq. (\ref{eq:minimumbitrateeq}).
\subsection{CNeT Context Model}
\label{subsec:contextmodel}

\begin{figure}
\centering
\captionsetup{justification=justified}
\includegraphics[width=.55\linewidth]{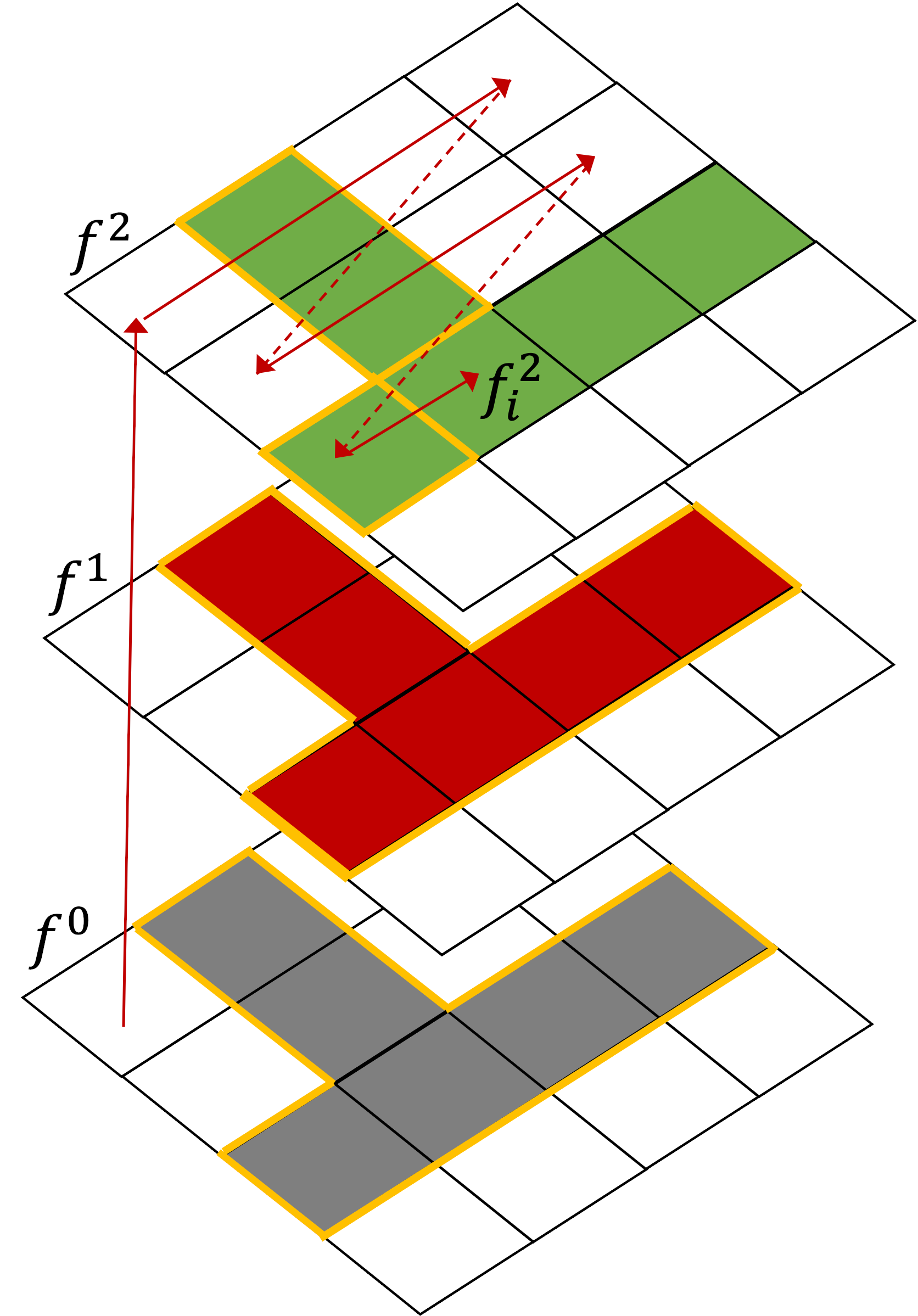}
\vspace{0.1cm}
\caption{An example of causal context in a ${4\times4}$  block. The bottom layer is occupancy feature, the next two layers are color features. The empty space is indicated with white color. The area inside the yellow boundaries are context to encode the current sub-point ${f^2_{\textcolor{black}{i}}}$. Red arrows show the 3D raster-scan order which is also the encoding order. \textcolor{black}{For visualization simplicity, we do not show the prediction of ${f^3}$.}}
\label{fig:3Dcontext}
\end{figure}

\par As previously discussed, in this paper our focus is lossless compression of features $F$ containing four different components $F=\{f^0, f^1, f^2, f^3\}$. Let $p(F)$ be the joint distribution of all feature components and all we need is to approximate $p(F)$. We can factorize $p(F)$ as a product of conditional distributions over the feature dimension as follows: 
\begin{equation} \label{eq:factorization}
\begin{split}
   p(F)=p(f^0) \cdot &p(f^1|f^0) 
   \cdot p(f^2|f^1,f^0) \cdot p(f^3|f^2,f^1,f^0)
\end{split}
\end{equation}
The conditional distribution of the occupancy feature can be further factorized in the spatial dimension as: 
\begin{equation} \label{eq:occfactorization}
\begin{split}
   p(f^0)&=\prod_{i=1}^{d^3}p(f_{i}^0|f_{i-1}^0,f_{i-2}^0,\ldots,f_{1}^0).
\end{split}
\end{equation}
with $d$ being the resolution of the voxelized point cloud and  $ i\in [1,d^3]$ the index voxels contained in the $d\times d\times d$ block, similar to \cite{nguyen2022learning}. In contrast, the last three terms in Eq. (\ref{eq:factorization}) are conditioned on the occupancy feature, which means the occupancy state of a voxel is known. Hence, the color feature distributions can be factorized as a product of $N$ sub-point conditional distributions. The equation below shows the full factorization for the third feature:
\begin{equation} \label{eq:subfactorize}
\begin{split}
   p(f^2|f^1,f^0)&=\prod_{i=1}^{N}p(f^1_i| f^1_{\textcolor{black}{i}-1},f^1_{\textcolor{black}{i}-2},\ldots,f^1_{1},f^1, f^0).
\end{split}
\end{equation}
where $N$ is the number of occupied voxels/points which is significantly smaller than $d^3$ as most voxels in the $d\times d\times d$ space are empty. The indexing is performed in 3D raster-scan order as for the occupancy feature but we skip empty voxels. Each term on Eq. (\ref{eq:subfactorize}) is the probability distribution of sub-point feature $f^1_i$ given all previous sub-points referred to as causal context as illustrated in Figure \ref{fig:3Dcontext}. 
\par We model each sub-point conditional distribution using a neural network which we call CNeT (Context NeTwork). To encode $F$, we can consider each sub-point (e.g. $f^0_i$) as a symbol stream and encode this $i^{th}$ symbol to a bitstream using $\hat{p}(f^0_i|f^0_{i-1},f^0_{i-2},...,f^0_{1})$, which is an estimation of $p(f^0_i|f^0_{i-1},f^0_{i-2},...,f^0_{1})$. As a result, this introduces a causal condition on both feature and spatial dimension for the encoding order as well as our networks.

\begin{figure*}
\centering
\captionsetup{justification=justified}
\includegraphics[width=.95\linewidth]{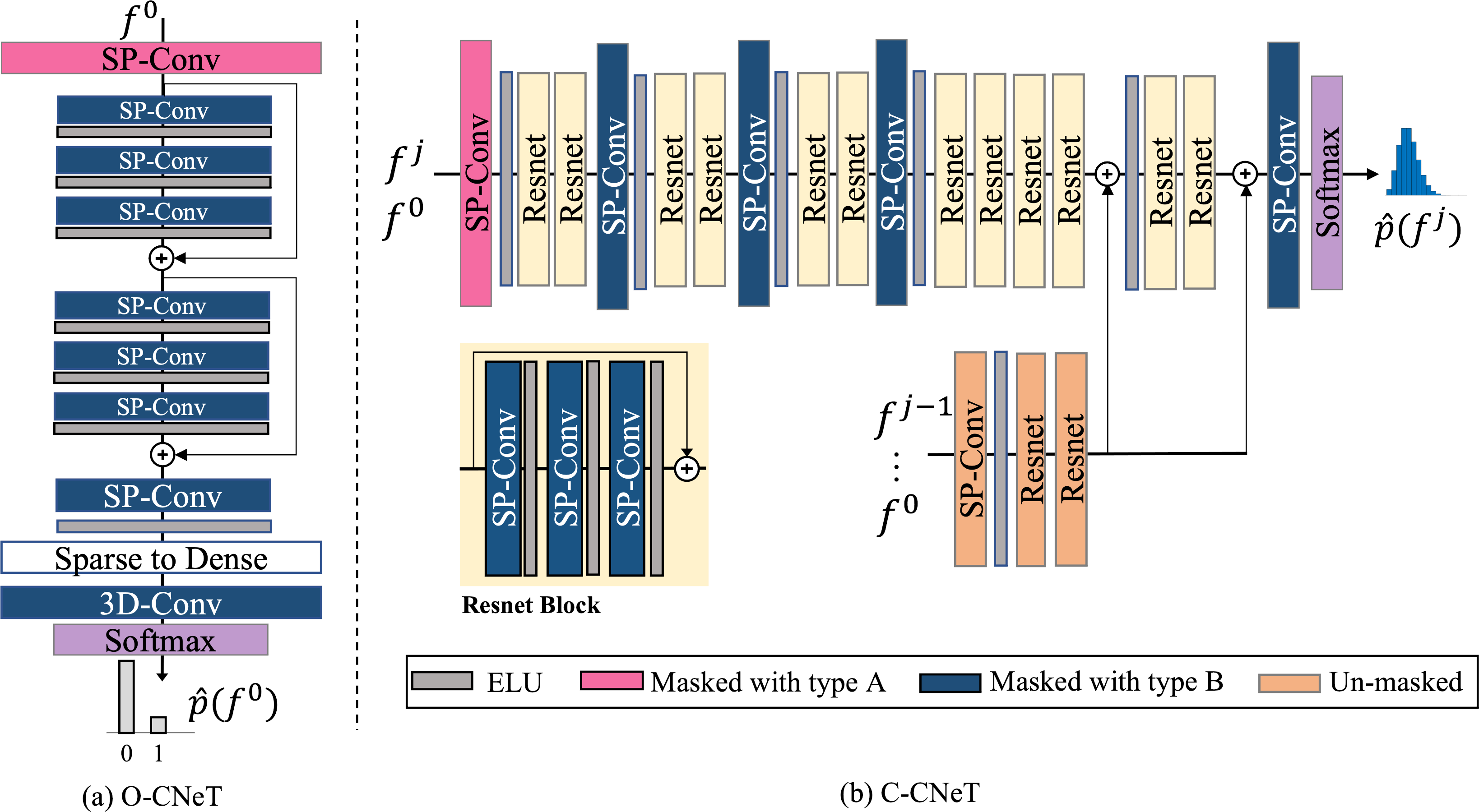}
\vspace{0.1cm}
\caption{CNeT neural network architectures. (a): Occupancy CNeT (O-CNeT), (b): Color feature CNeT (C-CNeT). Masks are applied in all convolution layers in both networks except for the previous feature path colored in bright brown. Type A masks are applied in the first layer of O-CNeT and C-CNeTs while type B masks are applied in the subsequent layers including Resnet block. In the Resnet blocks, the middle convolution layers has $kernel=3$ while the two other layers have $kernel=1$.}
\label{fig:network}
\end{figure*}

\subsection{Network Architecture}
\par In our implementation, we model the probability distribution of features using separate neural networks, using the context model scheme as discussed in Section \ref{subsec:contextmodel}. 
\begin{algorithm2e}[h]
\SetAlgoLined
\SetKwInOut{Input}{Input}
\KwIn{$mask\_type$} 
$kernel\_size \leftarrow \textcolor{black}{k_Dk_Hk_W}$ \\
$mask \leftarrow$ $ ones\_like(kernel)$\\
$mask\left[kernel\_size//2+ mask\_type==B,:,: \right]=0$\\
$kernel=kernel*mask$
       
\caption{Sparse mask construction and application}
\label{algo:maskalg}
\end{algorithm2e}
\par The context neural network for occupancy feature (O-CNeT) is shown in Figure \ref{fig:network}(a). Similar to \cite{nguyen2022learning}, to enforce the causality constraints,  we apply a sparse type A mask on the first layer and a sparse type B mask on the subsequent layers. The sparse type A mask restricts the connection from the current point to its feature and all the future points while the sparse type B mask allows the connection from the current spatial location to itself. The type A and type B sparse mask construction is presented in Algorithm \ref{algo:maskalg}. The first layer also plays the role of a coordinate generator which forces the availability of the next coordinate in 3D raster-scan order. Next, we employ two residual blocks before converting a point cloud to a voxel representation in the “Sparse to Dense” layer. The last convolution layer is a conventional 3D convolution with a kernel size of 1 and thus does not violate the causality constraints. The output is the probability of each voxel being occupied or empty, thus having the size of $d \times d \times d \times 2$.
\par The network architectures for color features (C-CNeT) are shown in Figure \ref{fig:network}(b). The probability distributions of sub-points are conditioned on the previous sub-points on the same color dimension and the previous features. Thus, there are two input branches, the main branch connects each sub-point to its previous points while the second branch connects them to the previous features. In the first branch, we apply a masking scheme similar to the O-CNeT to enforce the causality. Specifically, a type-A mask is applied in the first layer while type B masks are applied in the subsequent layers. The first branch is a sequence of ten residual blocks alternated by convolution layers. In contrast, there is no mask in the second branch, as the previous feature is allowed to connect to the output without any restriction. We merge two branches by an addition followed by masked layers at the very end of the network to produce the joint conditional probability distributions while avoiding losing information from the second branch under masking layers. The ELU activation function is applied after each convolutional layer except the last layer with soft-max activation. Note that our network supports different bit depths between features. For example, for the 8-bit depth attributes (R, G, B, Y), 256-way softmax is used, but for 9-bit depth attributes (Co, Cg) we replace this by a 512-way softmax activation. Bit depth difference will make no change to the input layer as all features will be normalized into $\left[-1,1\right]$ range before passing to the network. In the last layer, we only extract the feature $F_l$  containing the color probability distribution at each point from the tensor $\{ C, F_l\}$ as the output. 
\par Recall from Subsection \ref{overview} that our goal is to minimize the bit cost needed to encode the feature $F$ by minimizing the lower bound of the expected bits per symbol $H(\hat{p},p)$, where $\hat{p}$ is predicted discrete softmax distribution of $F$ . However, as we estimate the probability distribution model of each feature using separate neural networks, the total loss consists of four cross-entropy loss components corresponding to four features:

\begin{equation} \label{eq:lossfunction}
\begin{split}
   Loss &= H(\hat{p},p)=H(\hat{p}_{f^0},p_{f^0}) + H(\hat{p}_{f^1},p_{f^1})  
   \\
   &+H(\hat{p}_{f^2},p_{f^2}) +H(\hat{p}_{f^3},p_{f^3}) 
\end{split}
\end{equation}
\subsection{Point Cloud Encoder}
\par Prior to the actual encoding process, we first partition the point cloud into non-overlapping occupied blocks of size 64 and use an octree to signal the partitioning. The architecture of our encoder is shown in Figure \ref{fig:system overview}.  We encode four features sequentially in the $f^0,f^1,f^2,f^3$ order so that the previously encoded features are used as the additional context to encode the current feature. Within each feature, we encode sub-point features sequentially from the first to the last point in a generative manner. Every time a sub-point feature is encoded, it is fed back into our networks to predict the probability of the next sub-point except for the first sub-point where we use a uniform distribution. The discrete probability distribution is then passed to an arithmetic coder to encode the feature to the bitstream. To speed-up the running time, CNeT codec can process up to 8 blocks simultaneously. In the end, our bitstream consists of six components: metadata including the bit-depth, the unit coding block size; octree partition signaling; and the bitstreams of four feature components.
\subsection{Color Space Conversion}

\begin{figure}
\centering
\captionsetup{justification=justified}
\includegraphics[width=.99\linewidth]{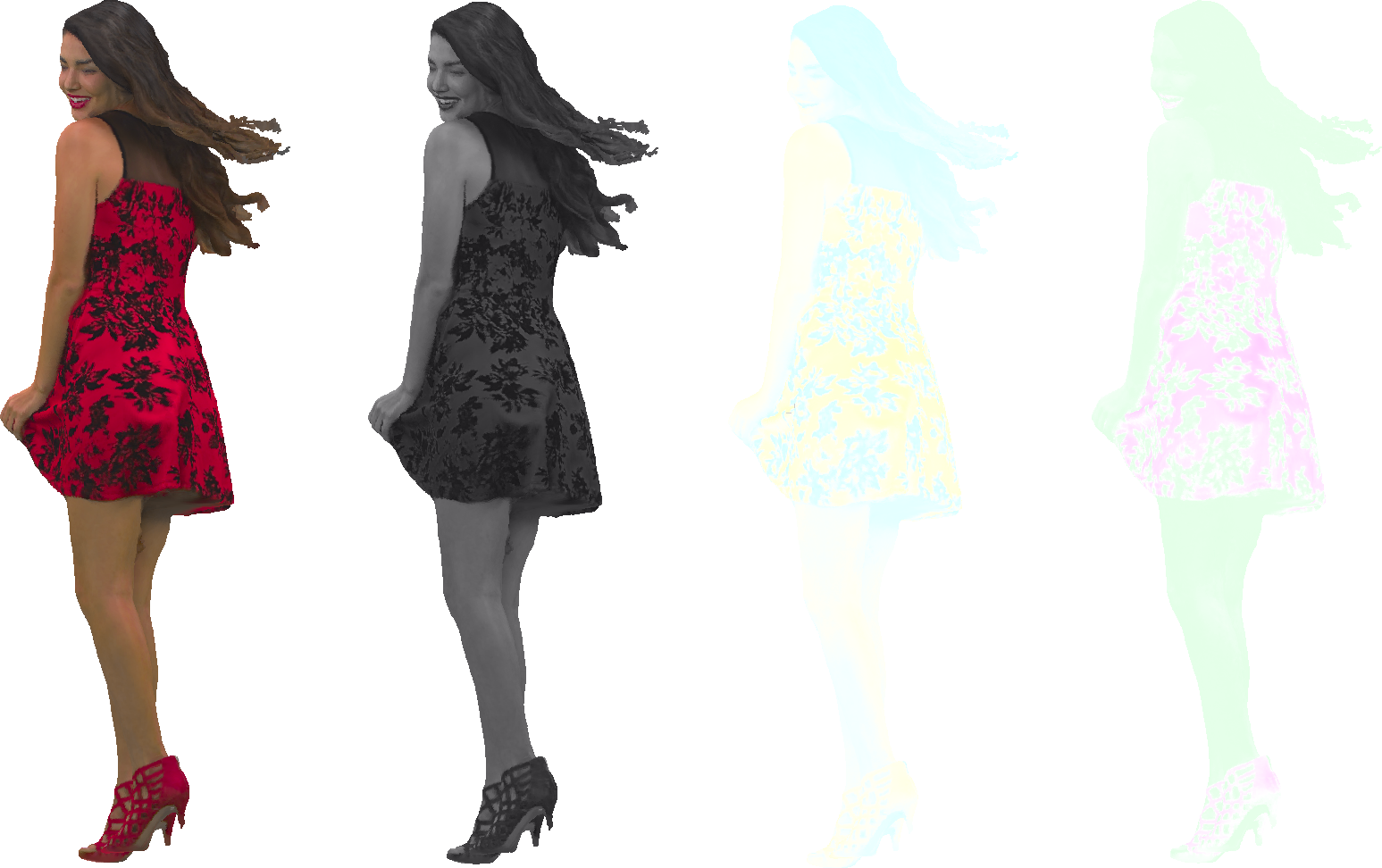}
\vspace{0.1cm}
\caption{From left to right: original Red$\&$Black point cloud from MPEG \cite{8i} on RGB color space and the representation of the individual color component Y, chrominance orange Co and chrominance green Cg.}
\label{fig:RGBtoluma}
\end{figure}

\par The RGB color space is usually used as the straight forward option for point cloud coding \cite{huang2021hierarchical,wang2022sparse} and color probabilistic distribution modeling \cite{oord2016pixel, salimans2017pixelcnn++}. However, luminance-chrominance color spaces have been proven to be more efficient for coding than RGB \cite{graziosi2020overview,sullivan2012overview}. In this paper, besides directly encoding the RGB color features, we also indirectly encode color features in the luminance-chrominance color space. Specifically, we adopt the lossless transformation from RGB to YCoCg as proposed in \cite{yuvtransform} and encode the Y, Co, Cg feature. On the decoder side, after decoding the Y, Co, and Cg, we losslessly perform an inverse transform to obtain the original RGB color. We believe that, by firstly encoding luminance Y, which contains most of point cloud detail (as illustrated in Figure \ref{fig:RGBtoluma}), and then using it to model the chrominances, we can efficiently exploit the correlation between channels and thus minimize the bit cost. In terms of implementation, the probabilistic distribution model proposed in \ref{subsec:contextmodel} is still the same. However, as the chrominance features will result in 9 bits depth, the final layer in the network for Co and Cg is changed from 256-way softmax to 512-way softmax.

\section{Experimental Results}
\label{experiment}
\subsection{Experimental Setup}

\begin{table}[t]
\caption{Number of blocks in the training and valid sets}
\centering
\resizebox{0.99\linewidth}{!}{
\begin{tabular}{|l|c|c|c|c|c|}
\cline{2-6}
\multicolumn{1}{c|}{}
&\begin{bf}MVUB\end{bf}
& \begin{bf}8i\end{bf} 
&\begin{bf}CAT1\end{bf}
& \begin{bf}Owlii\end{bf} 
& \begin{bf}Total\end{bf} \\
\hline
Training & 33000 &33000 & 9396 & 10000 & 85396\\
\hline
Validation & 5000 &5000 & 415 & 5000 & 15415\\
\hline
\end{tabular}}
\label{table:trainset}
\end{table}
\textbf{Training Data} Following the setting of SparseVoxelDNN \cite{nguyen2022learning}, we consider point clouds from four datasets: Microsoft Voxelized Upper Bodies (MVUB) \cite{loop2016microsoft} - dynamic dense voxelized upper-body point clouds, MPEG Owlii \cite{d20178i} and 8i \cite{8i}  - dynamic dense voxelized full-body point clouds and MPEG CAT1 \cite{8i} - static sparse point clouds for cultural heritage. Point clouds from CAT1 are sampled to 10-bit precision as in MVUB, Owlii, and 8i. We partition all point clouds to blocks of size 64 and randomly select a subset from three large datasets (MVUB, MPEG 8i, and Owlii) to avoid long training time. Table \ref{table:trainset} reports the number of blocks from each datasets.

\textbf{Data Augmentation} We apply a sub-sampling data augmentation technique in the training pipeline to improve the generalization of the networks, especially when our test set is a diverse set of sparse and dense point clouds. Besides, to train the Y, Co, Cg models, we perform online transformation from the original RGB color space. We normalize all inputs to the range $\left[-1, 1\right]$ before applying neural networks as shown in Eq. (\ref{eq:normalization}) with $bitdepth$ being the highest bit-depth in the input (e.g. 8 for RGB and 9 for YCoCg). 
\\
\begin{equation}
\label{eq:normalization}
\centering
\begin{split}
s&=\dfrac{2^{bitdepth}-1}{2}\\
x'&=\dfrac{x-s}{s}
\end{split}
\end{equation}
\par Due to the different bit-depth between YCoCg features, the Y feature will be ended up in $\left[-1, 0\right]$, hence, we shift Y' to the symmetrical range $\left[-0.5, 0.5\right]$ by an addition. Sub-sampling, color space transformation as well as normalization are implemented and embedded directly into our input pipeline while training. However, due to the randomness of the sub-sampling technique, we do not conduct online sub-sampling on the validation set but only sample once and use this in all validation steps.
\\
\par \textbf{Test Data} Our test set consists of 11 point clouds at 10 bit-depth from four different datasets: MVUB, MPEG 8i, MPEG CAT1, 8i and USP \cite{usp}.  All selected point clouds are either used in the MPEG Common Test Condition or the JPEG Pleno Common Test Condition to evaluate point cloud coding schemes. \textcolor{black}{The first two datasets are composed of two full-body voxelized dense point clouds (Phil\_10 and Ricardo\_76) from MVUB \cite{loop2016microsoft}  and four upper-body dense voxelized point clouds from MPEG 8i dataset \cite{8i, d20178i} (red$\&$black\_vox10\_1550, loot\_vox10\_1200, thaidancer\_vox10, boxer\_vox10), which is the same set from the recently introduced methods \cite{fu2022octattention, kaya2021neural,wang2022sparse, nguyen2021lossless, nguyen2021multiscale}. Besides, we additionally add point clouds for cultural heritage applications from the MPEG CAT1 dataset and the University of São Paulo dataset to create a diverse test set. The point clouds from MPEG CAT1 and USP are sparser and contain rougher surfaces than those from MVUB and MPEG 8i \cite{nguyen2021lossless}.} The test dataset is diverse in terms of sparsity, content type and capturing method, further details of the test point clouds can be found in \cite{nguyen2021lossless}. Note that all the point clouds in the test set have not been used in the training phase.
\\
\par \textbf{Training Strategy} We implement our models in PyTorch powered by Minkowski Engine \cite{choy20194d} with sparse convolution. Training and testing are performed on Xeon(R) W-1390 CPU and one GeForce RTX 3090 GPU (24 GB memory). We use batch sizes of 32, maximum epochs of 150 and an Adam optimizer. We found that applying a learning rate schedule with $step=2$, $gamma=0.95$ and the starting learning rate of $15e-5$ provides the optimal training result.

 \subsection{Performance Evaluation}
 \label{performanceevaluation}
\begin{center}
\centering
\begin{table*}[ht]
\caption{Average rate in bpp of the proposed method and percentage gains compared with MPEG G-PCC v14 (negative percentages denote a bitrate reduction).}
\resizebox{0.99\linewidth}{!}{ \begin{tabular}{|P{0.80cm}|l|R{0.9cm}|R{0.9cm}|R{0.9cm}|R{0.9cm}|R{0.9cm}|R{0.9cm}| R{1.1cm}|R{1.1cm}|R{1.1cm}| 
}
\cline{3-11}
\multicolumn{2}{c|}{}

& \multicolumn{3}{c|}{\begin{bf} G-PCC \end{bf}}
& \multicolumn{3}{c|}{\begin{bf} CNeT \end{bf}}
& \multicolumn{3}{c|}{\begin{bf} Gain over G-PCC \end{bf}}
\\
\hline
Dataset&Point Cloud&Geo&Color&Total&Geo&Color&Total &Geo&Color&Total \\
\hline

\multirow{3}{*}{MVUB}&Phil& 1.15& 10.25&	11.40&	0.40	&4.77	&5.17&	-65.49\% &-53.46\% &-54.68\% \\ 
\cline{2-11}
&Ricardo & 1.07	&5.88&	6.95	&0.35	&2.61	&2.96&	-67.29\%	&-55.61\% &-57.41\%\\
\cline{2-11}
&\textbf{Average} &\textbf{1.11}&	\textbf{8.07}	&\textbf{9.17}&	\textbf{0.37}	&\textbf{3.69}	&\textbf{4.06}	&\textbf{-66.39}\%
&	\textbf{-54.54\%}
	& \textbf{-56.04\%} \\
\cline{2-11}
\hline

\multirow{5}{*}{8i}&Red\&black &1.09&	9.33&	10.42	&0.39	&7.21	&7.60	&-64.22\%	&-22.72\%	&-27.06\% \\
\cline{2-11}
&Loot &0.95&6.35&7.30&0.33&4.65&4.98&-65.26\% & -26.77\% & -31.78\%\\
\cline{2-11}
&Thaidancer &1.00&10.84&11.84&0.32&9.49&9.81&-68.00\% & -12.45\% & -17.15\%\\
\cline{2-11}
&Boxer&0.95&6.95&7.90&0.30&5.73&6.03&-68.42\% & -17.55\% & -23.67\% \\
\cline{2-11}
&\textbf{Average} &\textbf{1.00}&\textbf{8.37}&\textbf{9.37}&\textbf{0.34}&\textbf{6.77}&\textbf{7.11}&\textbf{-66.48\%} &\textbf{ -19.88\%} &\textbf{ -24.92}\%\\
\hline

\multirow{4}{*}{CAT1}&Frog& 1.92&9.90&11.82&1.23&9.63&10.86&-35.95\% & -2.73\% & -8.12\%\\
\cline{2-11}
&Arco Valentino &4.85&15.89&20.74&3.03&15.76&18.79&-37.53\% & -0.82\% & -9.40\%  \\
\cline{2-11}
&Shiva &3.67&15.30&18.97&2.66&15.10&17.76&-27.52\% & -1.31\% & -6.38\% \\
\cline{2-11}
&\textbf{Average} &\textbf{3.48}&\textbf{13.70}&\textbf{17.18}&\textbf{2.31}&\textbf{13.50}&\textbf{15.80}&\textbf{-33.66\%} & \textbf{-1.62\%} & \textbf{-7.97\%}\\
\hline
\multirow{3}{*}{USP}&BumbaMeuBoi &
2.31&10.86&13.17&1.81&10.14&11.95&-21.65\% & -6.63\% & -9.26\% \\
\cline{2-11}
&RomanOiLight &2.43&13.45&15.88&1.98&13.28&15.26&-18.52\% & -1.26\% & -3.90\% \\
\cline{2-11}
&\textbf{Average}  & \textbf{2.37}&\textbf{12.16}&\textbf{14.53}&\textbf{1.90}&\textbf{11.71}&\textbf{13.61}&\textbf{-20.08\%} & \textbf{-3.95\% }&\textbf{ -6.58\%}\\
\hline
\hline
 \multicolumn{2}{|c|}{\textbf{AVERAGE} } & \textbf{1.94}&	\textbf{10.45}&	\textbf{12.40}&\textbf{	1.16}&\textbf{	8.94}&\textbf{	10.11}&\textbf{	-49.08\%}&\textbf{	-18.30\%}&\textbf{	-22.62\%}\\
\cline{2-11}
\hline
\hline
\end{tabular}}
\label{table:result table}
\end{table*}
\end{center}

\begin{figure}
\captionsetup{singlelinecheck = false, justification=justified, font=small, labelsep=space}
\begin{minipage}[b]{.485\linewidth}
  \centering
  \centerline{\includegraphics[width=0.99\linewidth]{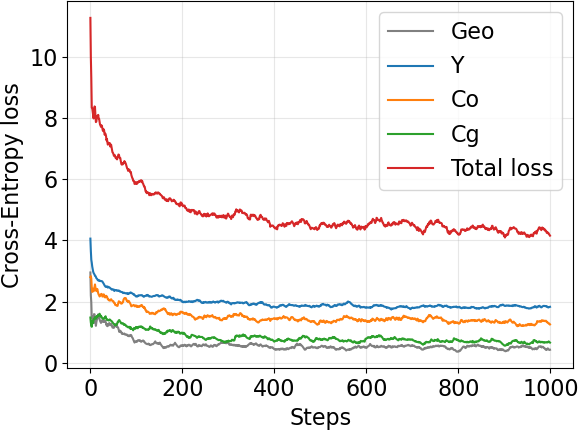}}
  \centerline{(a)}\medskip
\end{minipage}
\hfill
\begin{minipage}[b]{0.49\linewidth}
\label{sfig:typeA}
  \centering
  \centerline{\includegraphics[width=0.99\linewidth]{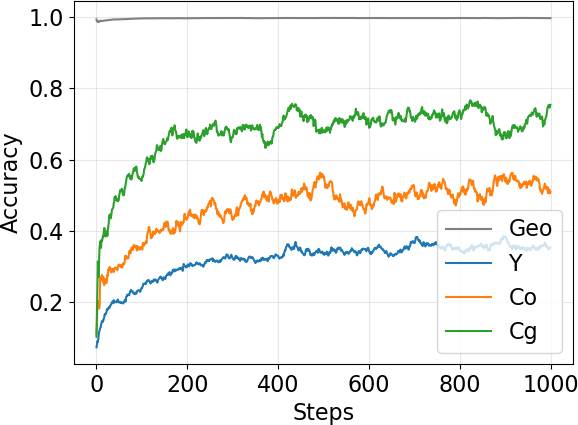}}
  \centerline{(b)}\medskip
\end{minipage}

\hfill

\caption{Loss and accuracy for Geometry, Y, Co, Cg features in training. The training steps are scaled to 0-1000 range. }
\label{fig:trainingperformance}
\end{figure}

\textbf{Training Performance} We train all models until they converge, the training time for O-CNeT is two days while C-CNeT models take seven days to the convergence. To evaluate the training performance, we track the loss of all four CNeT models. Besides, we cast the probabilistic distribution prediction problem to a binary/multi-class classification problem by selecting the highest score from the softmax distribution to obtain the accuracy metric. The losses and accuracies for Geometry, Y, Co, Cg CNeT models are illustrated in Fig \ref{fig:trainingperformance}. Specifically, in Fig \ref{fig:trainingperformance}(a) we report the loss for each feature and the total loss of all features. The total loss corresponds to the expected lower bound of the average bit per point (bpp) and theoretically, the lower the total loss, the better the encoding performance. We obtain the final total loss of 4.31 bpp, and the component loss for Geometry, Y, Co, Cg are at 0.47, 1.87, 1.33, and 0.82 respectively.  In the accuracy plot at Fig \ref{fig:trainingperformance}, we obtain a higher accurracy on Cg and Co features than Y feature. This is reasonable as the latter features are conditioned on the previous features. In addition, from the information theory perspective, the luminance is spread over a larger spectrum band, with more uncertainty and thus having a higher entropy than chrominance features. The accuracy for the geometry feature is close to 1 from the very beginning of the training, however, this typically happens in imbalance classification problems as most of the voxels are empty.

\textbf{Compression Performance} Table \ref{table:result table} reports the encoding bitrate in bits per point (bpp) of CNeT and the lossless geometry-based compression method G-PCC version 14 from MPEG. We show the bpp for the geometry component and color attribute component as well as the total bitrate of each method. In the last three columns, the gains of CNeT over G-PCC are reported. The bottom row indicates the average number over all point clouds from four datasets while the average for each dataset is reported in the last row. 
\par First, we observe that CNeT obtains a significant rate reduction of $-49.08\%$ on the geometry feature compared to G-PCC lossless compression. The highest rate reduction at about $-66\%$ was obtained in MVUB and MPEG 8i datasets which are dense and smooth point cloud datasets. \textcolor{black}{ We obtain a lower rate reduction on sparse point clouds from MPEG CAT1 and USP datasets with roughly $-33\%$ and $-20\%$ on average bitrate reduction. This can be explained by the fact that the networks can efficiently learn the relations between points and predict more accurate probability on dense point clouds than on sparse point clouds.} 
\par Second, CNeT outperforms G-PCC on color compression for all test point clouds. We efficiently compress the color attribute with $-18.3\%$ of average rate reduction compared to the G-PCC color attribute compression module. In particular, CNeT saves $-54.54\%$, $-19.88\%$, $-1.31\%$, $-3.95\%$ color bitrate compared with G-PCC on MVUB, 8i, CAT1 and USP datasets, respectively. And as expected, we obtain a higher color coding gain on the first two datasets than on the last two datasets similar to geometry compression.  For point clouds with higher density, our CNeT model can produce more accurate probabilities, and thus benefit our coding bitrate. In contrast, the sparsity and noise make the prediction inaccurate. We alleviate the problem by having data augmentation in the training. In total, when combining the geometry and color attribute compression, CNeT obtains a significant bit rate reduction over G-PCC of about $-22.6\%$ on average. 

\begin{figure}
\centering
\captionsetup{justification=raggedright}
\includegraphics[width=.99\linewidth]{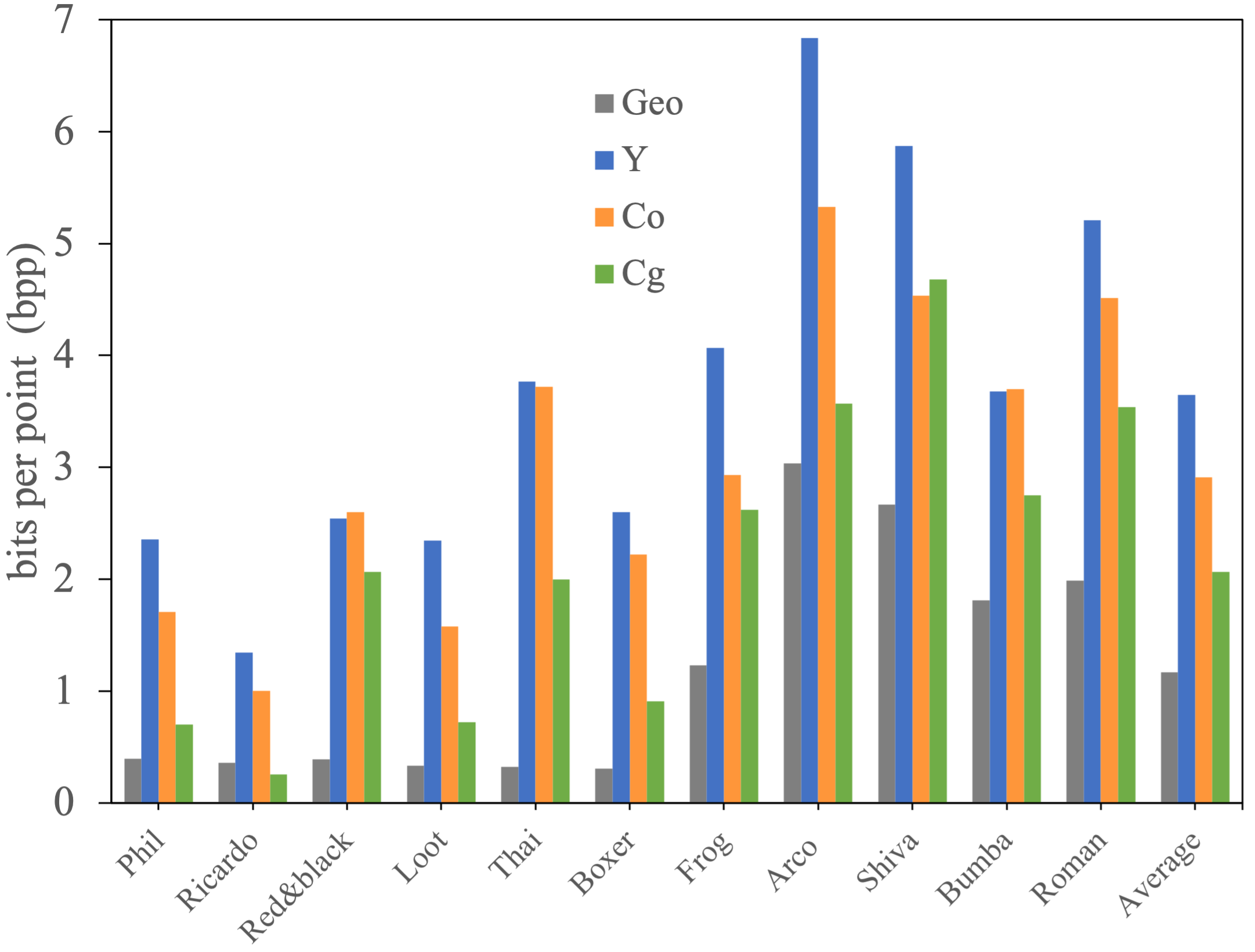}
\vspace{0.1cm}
\caption{Study on bit allocated for each feature.}
\label{fig:bitallocation}
\end{figure}

\textbf{Comparison to other Color Compression Methods} In addition to G-PCC, we compare CNeT method with a lossless compression method (HBDC) for static point clouds from \cite{huang2021hierarchical} and 3D-2D projection-based method V-PCC v18 \cite{8571288} from MPEG as shown in Table \ref{table:vshdbc}. Since the source code of the HBDC method is not available and HBDC is meant for attribute coding, we only report the available numbers for color attribute coding from the paper. We observe that both HBDC and V-PCC provide higher bitrates than G-PCC on all point clouds with $+22\%$ and $+25\%$ average bitrate increase, respectively. It should be noted that we use the higher version of V-PCC and G-PCC than the reference softwares used in the HBDC paper. Most importantly, CNeT significantly outperforms V-PCC, HBDC and G-PCC. The average color coding bitrate of CNeT is 4.81 bpp representing strong reductions of $39.0\%$ from G-PCC.  Therefore, we believe that neural network-based method is an promising avenue for future research on lossless point cloud attribute compression.
\begin{center}
\centering
\begin{table}[t]
\caption{Average bitrate in bpp for color compression and percentage gains of CNeT and other references compared
with MPEG G-PCC v14 \cite{8571288} (negative percentages denote a bitrate reduction).}
\resizebox{0.999\linewidth}{!}{ \begin{tabular}{|p{1.18cm}|wc{0.75cm}|wc{1.20cm}|wc{1.20cm}|wc{1.20cm}|
}

\cline{2-5}
\multicolumn{1}{c|}{}

& {\begin{bf} G-PCC \end{bf}}
& {\begin{bf} V-PCC \cite{8571288}\end{bf}}
& \begin{bf} HBDC \cite{huang2021hierarchical} \end{bf}
& {\begin{bf} CNeT \end{bf}}
\\
\hline
{{Phil}}&10.25&12.27 (+20\%)& 12.67 (+23\%)& 4.77 (-53\%)\\ 
\hline
{{Ricardo}}&5.88 & 6.96 (+18\%)&7.91 (+34\%)	&2.61 (-55\%)\\

\hline

{{Red\&black}}&9.33&11.57 (+24\%)& 11.39 (+22\%)&7.21 (-22\%)\\

\hline

{{Loot}}&6.35&8.10 (+27\%)& 7.79  (+22\%)&4.65 (-26\%)\\
\hline
\hline
{\textbf{AVERAGE} } & \textbf{7.95}&	\textbf{9.72 (+22\%)}&	\textbf{9.94 (+25\%)}&	\textbf{4.81 (-39\%)}\\
\cline{2-4}
\hline
\hline
\end{tabular}}
\label{table:vshdbc}
\end{table}
\end{center}

\textcolor{black}{\textbf{Comparison to other Geometry Compression Methods} Table \ref{table:geometry} compares the lossless geometry compression performance of the proposed method with state-of-the-art methods: SparsePCGC\cite{wang2022sparse}, VoxelDNN\cite{nguyen2021lossless}, NNOC\cite{kaya2021neural}, OctAttention\cite{fu2022octattention}. All methods were using similar training datasets as in CNeT except SparsePCGC.  Therefore, we additionally train from scratch our O-CNeT on dataset collection for AI-based point cloud experiments from MPEG \cite{mpegdataset} as in SparsePCGC \cite{wang2022sparse}. Specifically, we select two publicly accessible datasets: ShapeNet \cite{chang2015shapenet} with about 51.300 3D models and KITTI \cite{behley2019semantickitti} with 43,552 scans from 12 sequences of outdoor scenes for this training. We select the common test point clouds used by all methods for a fair comparison. From Table \ref{table:geometry}, we first notice a performance drop in the CNeT model trained on ShapeNet and KITTI datasets compare to the model trained on our training data. This can be explained by the fact that ShapeNet is a synthetic CAD model dataset and KITTI is a LIDAR sparse dataset while the test dataset composes of realistic and non-LIDAR point clouds. However, we observe that both CNeT and CNeT ShapeNet+KITTI offer significantly better gains than the other methods when compared with G-PCC. Thus we expect that a fully auto-regressive model such as CNeT can provide state-of-the-art bitrate reductions for lossless point cloud coding. }
\begin{center}
\centering

\begin{table*}[t]
\caption{Average bitrate in bpp for geometry compression and the percentage gains of CNeT and other references compared with MPEG G-PCC v14 (negative percentages denote a bitrate reduction). CNeT2 is trained on ShapeNet + KITTI datasets.}
\resizebox{0.99\linewidth}{!}{ 
\color{black}
\begin{tabular}{|l|c|c|c|c|c|c|c|}

\cline{2-8}
\multicolumn{1}{c|}{}

& \begin{bf} G-PCC  \end{bf}
& {\begin{bf} VoxelDNN  \end{bf}}
& {\begin{bf} NNOC  \end{bf}}
& {\begin{bf} OctAttention  \end{bf}}
& {\begin{bf} CNeT \end{bf}}
& {\begin{bf} SparsePCGC \end{bf}}
& {\begin{bf} CNeT2 \end{bf}}
\\
\hline
{{Red$\&$black }}&1.09&0.66 (-39\%)&0.73 (-33\%)&0.73 (-33\%)&0.39 (-64\%)&0.72 (-34\%) &0.52 (-53\%)\\ 
\hline
{{Loot}}&0.95&0.58 (-39\%)&0.58 (-38\%)&0.62 (-35\%)&0.33 (-65\%) &0.63 (-34\%) &0.46 (-52\%)\\ 

\hline

{{Thaidancer}}&1.00&0.68 (-39\%)&0.68 (-32\%)&0.65 (-35\%) &0.32 (-68\%) &0.67 (-33\%) &0.45 (-55\%)\\ 

\hline

{{Boxer}}&0.95&0.55 (-42\%)&0.55 (-42\%)&0.59 (-38\%) &0.30 (-68\%) &0.60 (-37\%) &0.45 (-53\%)\\ 
\hline
\hline
\textbf{AVERAGE} &\textbf{1.00}&\textbf{0.62 (-38\%)}&\textbf{0.64 (-36\%)}&\textbf{0.65 (-35\%)}&\textbf{0.34 (-66\%)}&\textbf{0.66 (-34\%)}  &\textbf{0.47 (-52\%)}\\ 
\hline
\hline
\end{tabular}
}
\label{table:geometry}
\end{table*}
\end{center}

\subsection{Ablation Study and Analysis}

\textbf{Bit Allocation} Figure \ref{fig:bitallocation}  illustrates the bpp allocated for each feature component which corresponds to the bit allocation in the bitstream. From this figure we can draw some observations. Geometry requires smaller space for storing than color components even though this is the first channel to be encoded. This can be explained by the fact that for each point there are only two possibilities of empty and non-empty, and thus the entropy of the source is smaller while the color components are having 256 or 512 possibilities and hence have more uncertainty. The color attribute consumes approximately $80\%$ of the bitstream and the latter the feature to be encoded, the less bit it consumes. This has already theoretically and practically been explained in the previous training result subsection.

\begin{center}
\centering
\begin{table}[t]
\caption{Average color rate in bpp of proposed method using RGB and YCoCg space. Percentage gains are compared with MPEG G-PCC v14 (negative percentages denote a bitrate reduction).}
\resizebox{1.\linewidth}{!}{ \begin{tabular}{|P{0.15cm}|l|wc{0.75cm}|wc{0.4cm}|wl{0.8cm}|wc{0.4cm}|wl{0.8cm}| }
\cline{3-7}
\multicolumn{2}{c|}{}

& \begin{bf} G-PCC \end{bf}
& \multicolumn{2}{c|}{\begin{bf} RGB \end{bf}}
& \multicolumn{2}{c|}{\begin{bf} YCoCg \end{bf}}
\\
\cline{2-7}
\multicolumn{1}{c|}{}&Point Cloud&bpp&bpp&Gain&bpp&Gain\\
\hline

\multirow{3}{*}{\rotatebox[origin=c]{90}{MVUB}}&Phil& 10.25&5.71&-44.29\%&4.77&-53.46\%	 \\ 
\cline{2-7}
&Ricardo & 5.88&3.09&-47.45\%&2.61&-55.61\%\\
\cline{2-7}
&\textbf{Average} &\textbf{8.07}&\textbf{4.40}&\textbf{-45.87\%}&\textbf{3.69}&\textbf{-54.54\%}	\\
\cline{2-7}
\hline

\multirow{5}{*}{\rotatebox[origin=c]{90}{8i}}&Red\&black &9.33&7.17&-23.15\%&7.21&-22.72\%	\\
\cline{2-7}
&Loot &6.35&4.59&-27.72\%&4.65&-26.77\%\\
\cline{2-7}
&Thaidancer &10.84&9.81&-9.50\%&9.49&-12.45\%\\
\cline{2-7}
&Boxer&6.95&5.74&-17.41\%&5.73&-17.55\% \\
\cline{2-7}
&\textbf{Average} &\textbf{8.37}&\textbf{6.83}&\textbf{-19.45\%}&\textbf{6.77}&\textbf{-19.88\%}\\
\hline

\multirow{4}{*}{\rotatebox[origin=c]{90}{CAT1}}&Frog&9.90&9.66&-2.42\%&9.63&-2.73\%\\
\cline{2-7}
&Arco Valentino &15.89&16.57&4.28\%&15.76&-0.82\% \\
\cline{2-7}
&Shiva &15.30&15.36&0.39\%&15.10&-1.31\% \\
\cline{2-7}
&\textbf{Average} &\textbf{13.70}&\textbf{13.86}&\textbf{0.75\%}&\textbf{13.50}&\textbf{-1.62\%}\\
\hline
\multirow{3}{*}{\rotatebox[origin=c]{90}{USP}}&BumbaMeuBoi &
10.86&10.82&-0.37\%&10.14&-6.63\%\\
\cline{2-7}
&RomanOiLight &13.45&13.46&0.07\%&13.28&-1.26\% \\
\cline{2-7}
&\textbf{Average}  &\textbf{12.16}&\textbf{12.14}&\textbf{-0.15\%}&\textbf{11.71}&\textbf{-3.95\%}\\
\hline
\hline

 \multicolumn{2}{|c|}{\textbf{AVERAGE} } & \textbf{10.45}&	\textbf{9.27}&	\textbf{-15.23\%}&\textbf{8.92}&\textbf{-18.30\%}\\
\cline{2-7}
\hline
\hline
\end{tabular}}
\label{table:colorspaceconversion}
\end{table}
\end{center}

\textbf{Color Space Conversion} In Table \ref{table:colorspaceconversion}, we report the bpp performance of encoding color features using direct RGB color space or the transformed YCoCg color space. Generally, we observe that direct coding on RGB space provides significant gain over G-PCC on dense point clouds. However, for Arco Valentino, Shiva, and RomanoOiLight point clouds, we obtain a slightly higher rate with $+4.28\%$, $+0.39\%$, $+0.07\%$ rate increase compared to G-PCC. However, we yield  $-15.23\%$ rate reduction on average compared to G-PCC with direct RGB coding.  By converting to YCoCg, we outperform G-PCC on all test point clouds with an average further bitrate reduction of  $-3.07\%$ over G-PCC corresponding to $-0.35$ bpp rate reduction compared to the RGB-based method. We expect that other attributes (e.g. reflectance, normal...) can also be efficiently compressed with an appropriate lossless transform.


\begin{figure}
\captionsetup{singlelinecheck = false, justification=justified}

\begin{minipage}[b]{0.48\linewidth}
\label{sfig:typeA}
  \centering
  \centerline{\includegraphics[width=0.99\linewidth]{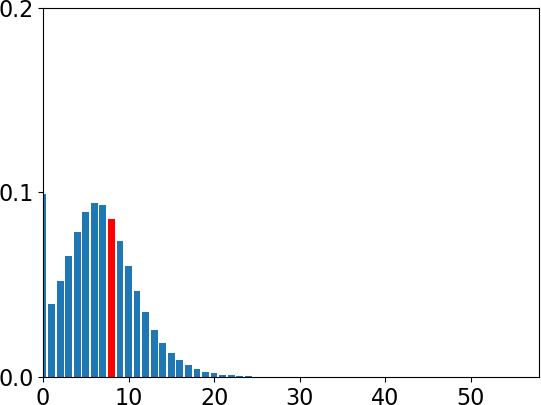}}
  \centerline{(a)}\medskip
\end{minipage}%
\hfill%
\begin{minipage}[b]{0.48\linewidth}
\label{sfig:typeA}
  \centering
  \centerline{\includegraphics[width=0.99\linewidth]{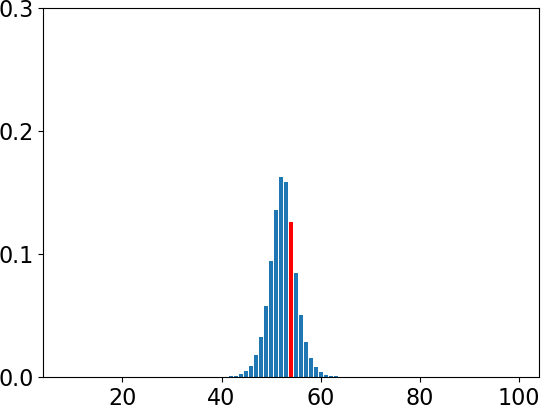}}
  \centerline{(b)}\medskip
\end{minipage}

\begin{minipage}[b]{.48\linewidth}
  \centering
  \centerline{\includegraphics[width=0.99\linewidth]{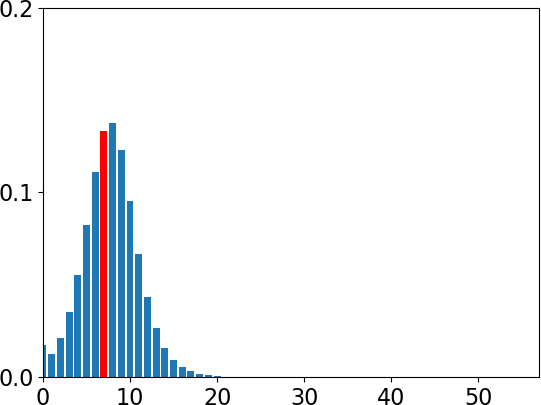}}
  \centerline{(c)}\medskip
\end{minipage}%
\hfill%
\begin{minipage}[b]{0.48\linewidth}
\label{sfig:typeA}
  \centering
  \centerline{\includegraphics[width=0.99\linewidth]{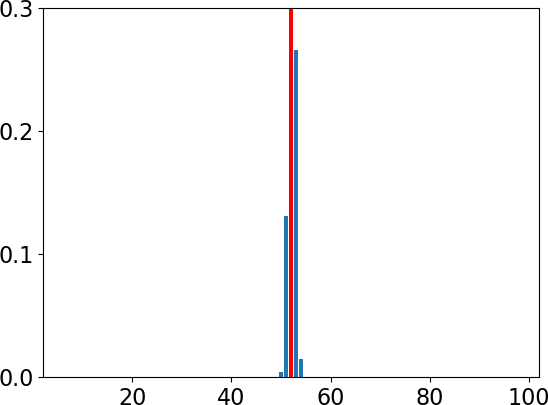}}
  \centerline{(d)}\medskip
\end{minipage}
\caption{Distributions produced by CNeT Softmax distribution and Mixture of Logistic distribution modeling. The red color indicates the probability of the ground truth. The x and y-axis are cropped for better visibility. (a), (b): Mixture of Logistic distribution, (c)(d): Softmax distribution.   }
\label{fig:distributon}
\end{figure}

\textbf{Probability Modeling} In contrast to the state-of-the-art probabilistic density estimation methods \cite{salimans2017pixelcnn++,mentzer2018conditional, theis2015generative} using Mixture of Logistic function (MoL) to model the continuous probability, we find the softmax discrete distribution to produce better performance in both training and encoding for our models. MoL produces symmetrical, smooth and meaningful distributions in terms of neighboring color categories as shown in Figure \ref{fig:distributon}(a)(b).  However, these are not necessary properties for the compression task. Besides, the values 0 and 255 usually get a much higher probability as they are the sum of the distribution lying outside the interval [0,255]  (as the example in Fig \ref{fig:distributon}(a)) which will impair the share of the true label. On the contrary, for the compression task, especially with a context-adaptive arithmetic coder, we would prefer having a high probability of the true color value regardless of the probability of other values or any properties of the distribution. Fig \ref{fig:distributon}(c) and (d) show the distribution from the Softmax-based model which are definitely more efficient than distributions generated from a MoL-based model in Fig \ref{fig:distributon}(a)(b). Note that the difference is negligible in binary classification problem with just two output probabilities. \\
\textbf{Aggregate Improvements} Table \ref{table:softmaxmol} compares the coding performance of YCoCg Softmax-based model and a baseline model with RGB color space and MoL on four datasets. Note that to obtain the MoL-based models, we replace the Softmax layer in the CNeT models by a mixture of logistic distributions. We find that a set of ten mixtures provides the best coding performance. From Table \ref{table:softmaxmol}, we observe that using Softmax discrete distribution plus color space transformation significantly reduces the coding bitrate compared to the baseline. Specifically, with the baseline model, we obtain $12.26$ bpp on average, a comparable rate to G-PCC at $12.40$ bpp. The YCoCg Softmax-based model produces lower rate at $10.11$ bpp on average and saves $-2.11$ bpp compared with the baseline. Note that the gain is paid by the training time as 256-way or 512-way Softmax makes gradients sparser with respect to the input than MoL with less output parameters. 

\begin{center}
\centering
\begin{table}[t]
\caption{Average rate in bpp of the Baseline (RGB+MoL) and the YCoCg Softmax-based model and the bpp difference (negative bpp denotes a bitrate reduction).}
\resizebox{0.99\linewidth}{!}{ \begin{tabular}{|l|c|c|c|
}

\cline{2-4}
\multicolumn{1}{c|}{}

& \begin{bf} Baseline \end{bf}
& {\begin{bf} YCoCg + Softmax \end{bf}}
& {\begin{bf} Gain \end{bf}}
\\
\hline
{{MVUB}}&6.60& 4.06&-2.54\\ 
\hline
{{8i}}&8.85 &7.11&-1.74	\\

\hline

{{CAT1}}&18.09&15.80&-2.28\\

\hline
{{USP}}&15.50&13.61&-1.89\\
\hline
\hline
{\textbf{AVERAGE} } & \textbf{12.26}&	\textbf{10.11}&	\textbf{-2.11}\\
\hline
\hline
\end{tabular}}
\label{table:softmaxmol}
\end{table}
\end{center}

\begin{figure}
\captionsetup{singlelinecheck = false, justification=justified}
\begin{minipage}[b]{.49\linewidth}
  \centering
  \centerline{\includegraphics[width=0.99\linewidth]{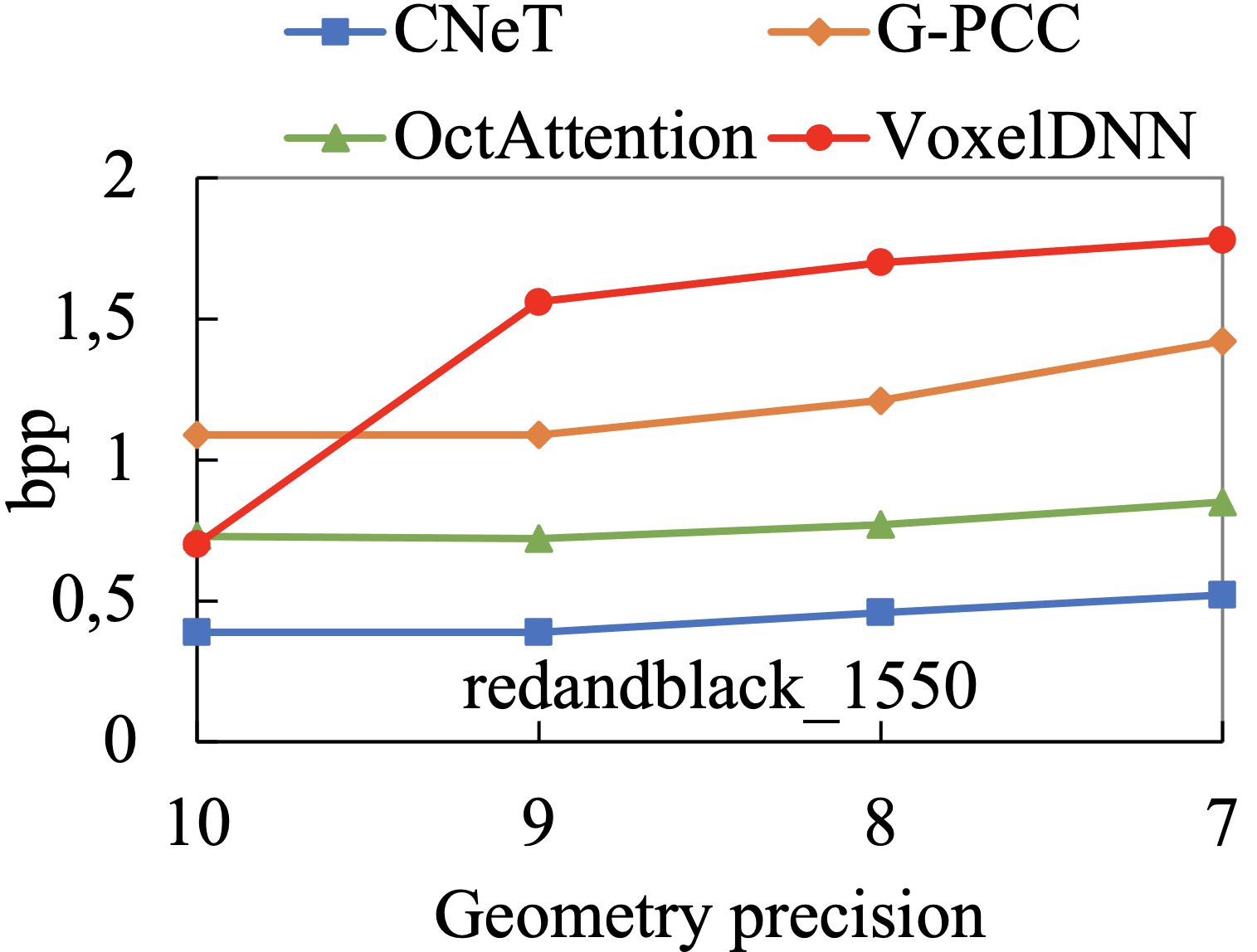}}
\end{minipage}
\begin{minipage}[b]{0.49\linewidth}
  \centering
  \centerline{\includegraphics[width=0.99\linewidth]{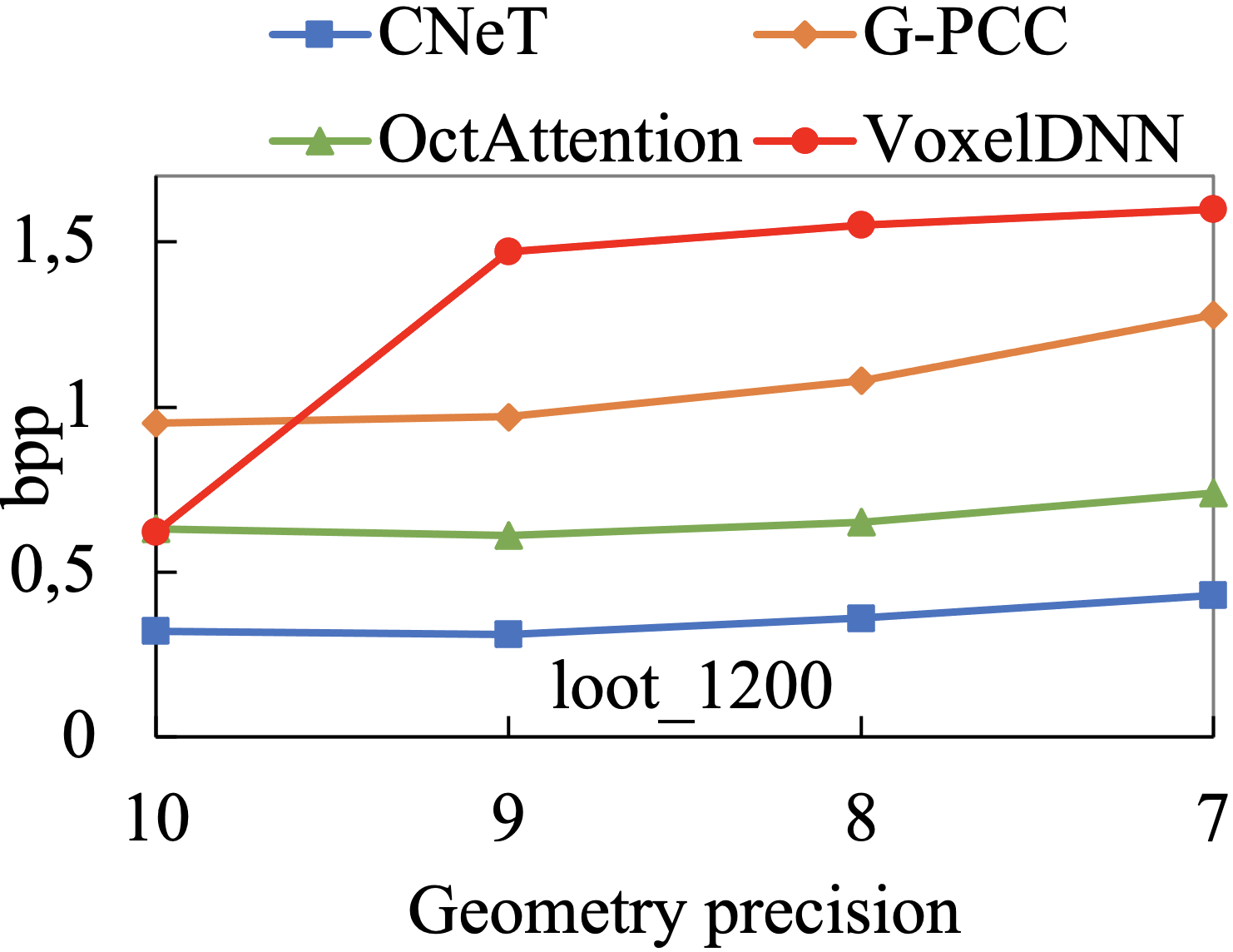}}
\end{minipage}

\begin{minipage}[b]{0.49\linewidth}
  \centering
  \centerline{\includegraphics[width=0.99\linewidth]{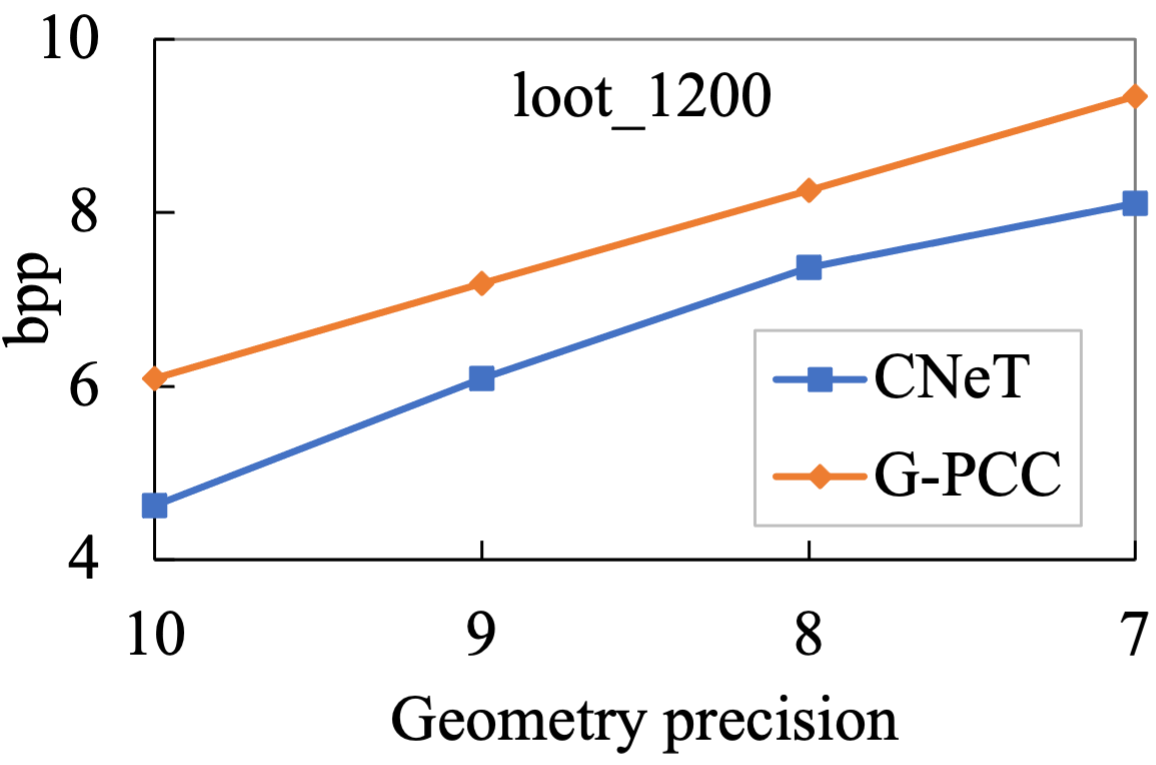}}
\end{minipage}
\begin{minipage}[b]{0.49\linewidth}
  \centering
  \centerline{\includegraphics[width=0.99\linewidth]{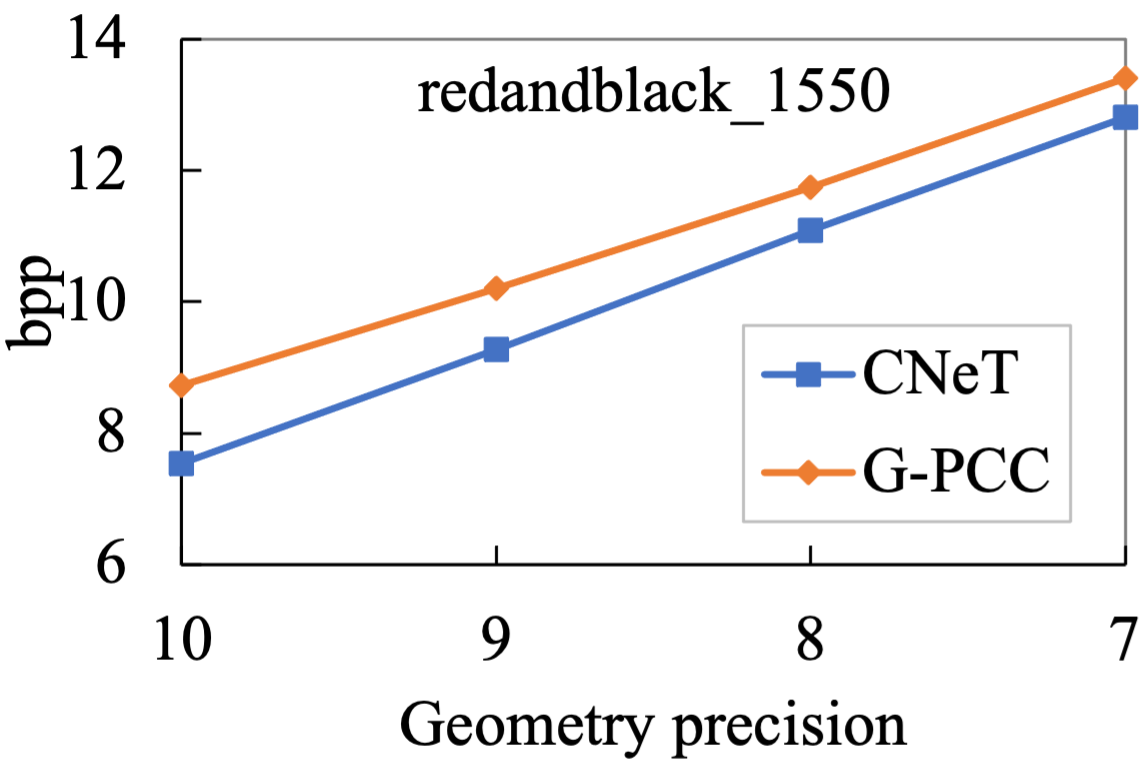}}
\end{minipage}
\hfill

\caption{\textcolor{black}{Performance comparison at different geometry precisions between CNeT and state-of-the-art methods. Top figures: only geometry compression. Bottom figures: geometry + attribute compression.}  }
\label{fig:geometrycmpare}
\end{figure}

\textbf{Performance at different geometry precision} \textcolor{black}{We perform additional experiments to evaluate the robustness of CNeT to density variation. We follow the setting in OctAttention \cite{fu2022octattention} to test our geometry and attribute compression at four geometry precision levels on $loot\_1200$, $red\_and\_black\_1550$ from the MPEG 8i dataset. 
The results are shown in Fig. \ref{fig:geometrycmpare}. Even though our bitrates increase due to lower point density while reducing the geometry bit-depth, we notice that our geometry and attribute compression method retain superiority over all competitors at all precision levels. These results show that CNeT is stable against various geometry precisions. }

\textbf{Complexity Analysis} \textcolor{black}{ In total, our models require less than 700 MB of disk storage space. We run our test on the same machine used for training, our model consumes only 5.4 GB GPU memory on average while encoding.} Table \ref{table:complexity} reports the running time of G-PCC and our codec on all test point clouds in seconds. Clearly, CNeT has a similar encoding time to G-PCC. However, CNeT decoding time is longer compared to G-PCC which is the cost paid for the significant performance improvement. This is a common drawback of all auto-regressive generative models such as PixelCNN \cite{oord2016pixel, salimans2017pixelcnn++} and VoxelDNN \cite{nguyen2021lossless}. At the encoder side we can perform parallel prediction, while at the decoder side we must predict the feature probabilities sequentially and thus having higher decoding times. In Table \ref{table:complexitygeo}, we report the running time of CNeT geometry coding and other geometry coding methods on the set of point clouds from Table \ref{table:geometry}. Since the source code of SparsePCGC \cite{wang2022sparse} is not available, we only report available running time from the paper. We observe that CNeT geometry encoding time is only behind octree-based methods (G-PCC and OctAttention \cite{fu2022octattention}) which are generally fast. We also notice that VoxelDNN \cite{nguyen2022learning}, NNOC \cite{kaya2021neural}, and OctAttention even have a longer decoding time than our method. The long decoding time of CNeT can be alleviated by multi-scale architectures or alternative approaches that avoid sequential probability estimation similar to VoxelDNN and PixelCNN. Besides, CNeT running time is highly dependent on the number of occupied blocks (few hundred blocks) that are disjoint and encoded independently. We expect that it is possible to further speed up the codec by utilizing parallelization or having better partitioning schemes to reduce the number of blocks.
%
%
\begin{table}[t]
\caption{Average runtime (in seconds) of CNeT encoder comparing with G-PCC v14.}
\centering
\resizebox{0.63\linewidth}{!}
{ \begin{tabular}{|M{1.0cm}|M{1.2cm}|M{1.5cm}|}
\cline{2-3}
\multicolumn{1}{c|}{}
&\begin{bf}G-PCC\end{bf}
& \begin{bf}CNeT\end{bf} 
\\
\hline
Enc &22&  25\\
\hline
Dec &12&  590\\
\hline
\end{tabular}}
\label{table:complexity}
\end{table}

\begin{table*}[tb]

\caption{Average runtime (in seconds) of CNeT geometry encoder comparing with other methods .}
\centering
\resizebox{0.63\linewidth}{!}
{
 \begin{tabu}{|M{0.5cm}|M{0.9cm}|M{1.1cm}|M{0.8cm}|M{1.4cm}|M{1.45cm}|M{0.8cm}|}

\cline{2-7}
\multicolumn{1}{c|}{}&\begin{bf}G-PCC\end{bf}
& \begin{bf}VoxelDNN\end{bf} 
& \begin{bf}NNOC\end{bf} 
& \begin{bf}OctAttention\end{bf}  
& \begin{bf}SparsePCGC\end{bf} 
& \begin{bf}CNeT\end{bf} \\
\hline
Enc &1.07&25&34&0.57&9.50&5.88\\
\hline
Dec &0.43&497 &341&573&9.10&235\\
\hline
\end{tabu}}

\label{table:complexitygeo}
\end{table*}

\section{Conclusions and future work}
\label{conclusion}
\par In this paper, we proposed the first learning-based lossless point cloud geometry and attribute compression scheme by employing a conditional probability model across the feature and point dimension. We build CNeT neural networks to efficiently exploit the correlation between points and accurately model the conditional probability for the arithmetic coder. We further propose using discrete softmax distribution and indirectly encoding color attributes in the YCoCg color space to minimize the coding bitrate. We evaluate our method on a set of sparse and dense point clouds from four different datasets. The results show that the proposed method achieves state-of-the-art \textcolor{black}{performance and robustness to various geometry precisions}. There are a number of potential directions for further exploration, including reducing the complexity with a multi-scale structure or developing a compression method for dynamic point clouds.

\ifCLASSOPTIONcaptionsoff
  \newpage
\fi

\bibliographystyle{./IEEEtran}
\bibliography{./IEEEabrv,./refs}

\begin{thebibliography}{10}
\providecommand{\url}[1]{#1}
\csname url@samestyle\endcsname
\providecommand{\newblock}{\relax}
\providecommand{\bibinfo}[2]{#2}
\providecommand{\BIBentrySTDinterwordspacing}{\spaceskip=0pt\relax}
\providecommand{\BIBentryALTinterwordstretchfactor}{4}
\providecommand{\BIBentryALTinterwordspacing}{\spaceskip=\fontdimen2\font plus
\BIBentryALTinterwordstretchfactor\fontdimen3\font minus
  \fontdimen4\font\relax}
\providecommand{\BIBforeignlanguage}[2]{{%
\expandafter\ifx\csname l@#1\endcsname\relax
\typeout{** WARNING: IEEEtran.bst: No hyphenation pattern has been}%
\typeout{** loaded for the language `#1'. Using the pattern for}%
\typeout{** the default language instead.}%
\else
\language=\csname l@#1\endcsname
\fi
#2}}
\providecommand{\BIBdecl}{\relax}
\BIBdecl

\bibitem{graziosi2020overview}
D.~Graziosi, O.~Nakagami, S.~Kuma, A.~Zaghetto, T.~Suzuki, and A.~Tabatabai,
  ``\BIBforeignlanguage{en}{An overview of ongoing point cloud compression
  standardization activities: video-based ({V}-{PCC}) and geometry-based
  ({G}-{PCC})},'' \emph{\BIBforeignlanguage{en}{APSIPA Transactions on Signal
  and Information Processing}}, vol.~9, 2020.

\bibitem{8571288}
\BIBentryALTinterwordspacing
S.~Schwarz, M.~Preda, V.~Baroncini, M.~Budagavi, P.~Cesar, P.~A. Chou, R.~A.
  Cohen, M.~Krivokuca, S.~Lasserre, Z.~Li, J.~Llach, K.~Mammou, R.~Mekuria,
  O.~Nakagami, E.~Siahaan, A.~Tabatabai, A.~M. Tourapis, and V.~Zakharchenko,
  ``\BIBforeignlanguage{en}{Emerging {MPEG} {Standards} for {Point} {Cloud}
  {Compression}},'' \emph{\BIBforeignlanguage{en}{IEEE Journal on Emerging and
  Selected Topics in Circuits and Systems}}, pp. 1--1, 2018. [Online].
  Available: \url{https://ieeexplore.ieee.org/document/8571288/}
\BIBentrySTDinterwordspacing

\bibitem{jang2019video}
E.~S. Jang, M.~Preda, K.~Mammou, A.~M. Tourapis, J.~Kim, D.~B. Graziosi,
  S.~Rhyu, and M.~Budagavi, ``Video-{Based} {Point}-{Cloud}-{Compression}
  {Standard} in {MPEG}: {From} {Evidence} {Collection} to {Committee} {Draft}
  [{Standards} in a {Nutshell}],'' \emph{IEEE Signal Processing Magazine},
  vol.~36, no.~3, pp. 118--123, May 2019.

\bibitem{8945224}
H.~Liu, H.~Yuan, Q.~Liu, J.~Hou, and J.~Liu, ``A comprehensive study and
  comparison of core technologies for mpeg 3-d point cloud compression,''
  \emph{IEEE Transactions on Broadcasting}, vol.~66, no.~3, pp. 701--717, 2020.

\bibitem{de2016compression}
R.~L. De~Queiroz and P.~A. Chou, ``Compression of 3d point clouds using a
  region-adaptive hierarchical transform,'' \emph{IEEE Transactions on Image
  Processing}, vol.~25, no.~8, pp. 3947--3956, 2016.

\bibitem{choy20194d}
C.~Choy, J.~Gwak, and S.~Savarese, ``4d spatio-temporal convnets: Minkowski
  convolutional neural networks,'' in \emph{Proceedings of the IEEE Conference
  on Computer Vision and Pattern Recognition}, 2019, pp. 3075--3084.

\bibitem{schnabel2006octree}
R.~Schnabel and R.~Klein, ``Octree-based point-cloud compression.''
  \emph{Spbg}, vol.~6, pp. 111--120, 2006.

\bibitem{garcia2017context}
D.~C. Garcia and R.~L. de~Queiroz, ``Context-based octree coding for
  point-cloud video,'' in \emph{2017 {IEEE} {International} {Conference} on
  {Image} {Processing} ({ICIP})}, Sep. 2017, pp. 1412--1416, iSSN: 2381-8549.

\bibitem{garcia2018intra}
D.~C. Garcia and R.~L.~d. Queiroz, ``Intra-{Frame} {Context}-{Based} {Octree}
  {Coding} for {Point}-{Cloud} {Geometry},'' in \emph{2018 25th {IEEE}
  {International} {Conference} on {Image} {Processing} ({ICIP})}, Oct. 2018,
  pp. 1807--1811.

\bibitem{garcia2019geometry}
D.~C. Garcia, T.~A. Fonseca, R.~U. Ferreira, and R.~L. de~Queiroz, ``Geometry
  {Coding} for {Dynamic} {Voxelized} {Point} {Clouds} {Using} {Octrees} and
  {Multiple} {Contexts},'' \emph{IEEE Transactions on Image Processing},
  vol.~29, pp. 313--322, 2019.

\bibitem{biswas2020muscle}
S.~Biswas, J.~Liu, K.~Wong, S.~Wang, and R.~Urtasun, ``Muscle: Multi sweep
  compression of lidar using deep entropy models,'' \emph{Advances in Neural
  Information Processing Systems}, vol.~33, 2020.

\bibitem{huang2020octsqueeze}
\BIBentryALTinterwordspacing
L.~Huang, S.~Wang, K.~Wong, J.~Liu, and R.~Urtasun, ``{OctSqueeze}:
  {Octree}-{Structured} {Entropy} {Model} for {LiDAR} {Compression},''
  \emph{arXiv:2005.07178 [cs, eess]}, May 2020. [Online]. Available:
  \url{http://arxiv.org/abs/2005.07178}
\BIBentrySTDinterwordspacing

\bibitem{dricot2019adaptive}
A.~Dricot and J.~Ascenso, ``Adaptive multi-level triangle soup for
  geometry-based point cloud coding,'' in \emph{2019 IEEE 21st International
  Workshop on Multimedia Signal Processing (MMSP)}.\hskip 1em plus 0.5em minus
  0.4em\relax IEEE, 2019, pp. 1--6.

\bibitem{neighbor}
``Neighbour-dependent entropy coding of occupancy patterns,'' in \emph{TMC3,
  ISO/IEC JTC1/SC29/WG11 input document m42238, Gwangju, Korea, January 2018.}

\bibitem{intracodinggpcc}
``Intra mode for geometry coding,'' in \emph{TMC3, ISO/IEC JTC1/SC29/WG11 input
  document m43600, Ljubljana, Slovenia, July 2018.}

\bibitem{planarcodingmode}
``Planar mode in octree-based geometry coding,'' in \emph{TISO/IEC
  JTC1/SC29/WG11 input document m48906, Gothenburg, Sweden, July 2019.}

\bibitem{angularcodingmode}
``An improvement of the planar coding mode,'' in \emph{ISO/IEC JTC1/SC29/WG11
  input document m50642, Geneva, CH, Oct 2019.}

\bibitem{kaya2021neural}
E.~C. Kaya and I.~Tabus, ``Neural network modeling of probabilities for coding
  the octree representation of point clouds,'' in \emph{2021 IEEE 23rd
  International Workshop on Multimedia Signal Processing (MMSP)}.\hskip 1em
  plus 0.5em minus 0.4em\relax IEEE, 2021, pp. 1--6.

\bibitem{fu2022octattention}
C.~Fu, G.~Li, R.~Song, W.~Gao, and S.~Liu, ``Octattention: Octree-based
  large-scale contexts model for point cloud compression,'' \emph{arXiv
  preprint arXiv:2202.06028}, 2022.

\bibitem{theis2017lossy}
L.~Theis, W.~Shi, A.~Cunningham, and F.~Husz{\'a}r, ``Lossy image compression
  with compressive autoencoders,'' \emph{arXiv preprint arXiv:1703.00395},
  2017.

\bibitem{quach2019learning}
M.~Quach, G.~Valenzise, and F.~Dufaux, ``Learning {Convolutional} {Transforms}
  for {Lossy} {Point} {Cloud} {Geometry} {Compression},'' in \emph{2019 {IEEE}
  {International} {Conference} on {Image} {Processing} ({ICIP})}, Sep. 2019,
  pp. 4320--4324, iSSN: 1522-4880.

\bibitem{yan2019deep}
W.~Yan, S.~Liu, T.~H. Li, Z.~Li, G.~Li \emph{et~al.}, ``Deep autoencoder-based
  lossy geometry compression for point clouds,'' \emph{arXiv preprint
  arXiv:1905.03691}, 2019.

\bibitem{wang2021multiscale}
J.~Wang, D.~Ding, Z.~Li, and Z.~Ma, ``Multiscale point cloud geometry
  compression,'' in \emph{2021 Data Compression Conference (DCC)}.\hskip 1em
  plus 0.5em minus 0.4em\relax IEEE, 2021, pp. 73--82.

\bibitem{wang2021sparse}
J.~Wang, D.~Ding, Z.~Li, X.~Feng, C.~Cao, and Z.~Ma, ``Sparse tensor-based
  multiscale representation for point cloud geometry compression,'' \emph{arXiv
  preprint arXiv:2111.10633}, 2021.

\bibitem{9287077}
M.~Quach, G.~Valenzise, and F.~Dufaux, ``Improved deep point cloud geometry
  compression,'' in \emph{2020 IEEE 22nd International Workshop on Multimedia
  Signal Processing (MMSP)}, 2020, pp. 1--6.

\bibitem{nguyen2021learning}
D.~T. Nguyen, M.~Quach, G.~Valenzise, and P.~Duhamel, ``Learning-based lossless
  compression of 3d point cloud geometry,'' in \emph{ICASSP 2021-2021 IEEE
  International Conference on Acoustics, Speech and Signal Processing
  (ICASSP)}.\hskip 1em plus 0.5em minus 0.4em\relax IEEE, 2021, pp. 4220--4224.

\bibitem{nguyen2021lossless}
------, ``Lossless coding of point cloud geometry using a deep generative
  model,'' \emph{IEEE Transactions on Circuits and Systems for Video
  Technology}, vol.~31, no.~12, pp. 4617--4629, 2021.

\bibitem{nguyen2021multiscale}
------, ``Multiscale deep context modeling for lossless point cloud geometry
  compression,'' in \emph{2021 IEEE International Conference on Multimedia \&
  Expo Workshops (ICMEW)}.\hskip 1em plus 0.5em minus 0.4em\relax IEEE, 2021,
  pp. 1--6.

\bibitem{nguyen2022learning}
D.~T. Nguyen and A.~Kaup, ``Learning-based lossless point cloud geometry coding
  using sparse representations,'' \emph{arXiv preprint arXiv:2204.05043}, 2022.

\bibitem{7025414}
C.~Zhang, D.~Florêncio, and C.~Loop, ``Point cloud attribute compression with
  graph transform,'' in \emph{2014 IEEE International Conference on Image
  Processing (ICIP)}, 2014, pp. 2066--2070.

\bibitem{7482691}
R.~L. de~Queiroz and P.~A. Chou, ``Compression of 3d point clouds using a
  region-adaptive hierarchical transform,'' \emph{IEEE Transactions on Image
  Processing}, vol.~25, no.~8, pp. 3947--3956, 2016.

\bibitem{8462684}
Y.~Xu, W.~Hu, S.~Wang, X.~Zhang, S.~Wang, S.~Ma, and W.~Gao, ``Cluster-based
  point cloud coding with normal weighted graph fourier transform,'' in
  \emph{2018 IEEE International Conference on Acoustics, Speech and Signal
  Processing (ICASSP)}, 2018, pp. 1753--1757.

\bibitem{4378368}
Y.~Huang, J.~Peng, C.-C.~J. Kuo, and M.~Gopi, ``A generic scheme for
  progressive point cloud coding,'' \emph{IEEE Transactions on Visualization
  and Computer Graphics}, vol.~14, no.~2, pp. 440--453, 2008.

\bibitem{gu20193d}
S.~Gu, J.~Hou, H.~Zeng, H.~Yuan, and K.-K. Ma, ``3d point cloud attribute
  compression using geometry-guided sparse representation,'' \emph{IEEE
  Transactions on Image Processing}, vol.~29, pp. 796--808, 2019.

\bibitem{TMC13}
\BIBentryALTinterwordspacing
M.~Group, ``Mpeg tmc13 reference software,'' (accessed May 02, 2022). [Online].
  Available: \url{https://github.com/MPEGGroup/mpeg-pcc-tmc13}
\BIBentrySTDinterwordspacing

\bibitem{lifting}
M.~K. T. A. K. J. R. F. V. V.~S. Y, ``Proposal for improved lossy compression
  in tmc1,'' in \emph{ISO/IEC JTC1/SC29/WG11 M42640, 2018}.

\bibitem{predicting}
M.~3DG, ``Pcc test model category 3 v0.'' in \emph{ISO/IEC JTC1/SC29/ WG11
  N17249, 2017.}

\bibitem{gu20203d}
S.~Gu, J.~Hou, H.~Zeng, and H.~Yuan, ``3d point cloud attribute compression via
  graph prediction,'' \emph{IEEE Signal Processing Letters}, vol.~27, pp.
  176--180, 2020.

\bibitem{liu2021hybrid}
H.~Liu, H.~Yuan, Q.~Liu, J.~Hou, H.~Zeng, and S.~Kwong, ``A hybrid compression
  framework for color attributes of static 3d point clouds,'' \emph{IEEE
  Transactions on Circuits and Systems for Video Technology}, vol.~32, no.~3,
  pp. 1564--1577, 2021.

\bibitem{7434610}
R.~Mekuria, K.~Blom, and P.~Cesar, ``Design, implementation, and evaluation of
  a point cloud codec for tele-immersive video,'' \emph{IEEE Transactions on
  Circuits and Systems for Video Technology}, vol.~27, no.~4, pp. 828--842,
  2017.

\bibitem{9169853}
L.~Li, Z.~Li, S.~Liu, and H.~Li, ``Efficient projected frame padding for
  video-based point cloud compression,'' \emph{IEEE Transactions on
  Multimedia}, vol.~23, pp. 2806--2819, 2021.

\bibitem{quach2020folding}
M.~Quach, G.~Valenzise, and F.~Dufaux, ``Folding-based compression of point
  cloud attributes,'' in \emph{2020 IEEE International Conference on Image
  Processing (ICIP)}.\hskip 1em plus 0.5em minus 0.4em\relax IEEE, 2020, pp.
  3309--3313.

\bibitem{alexiou2020towards}
E.~Alexiou, K.~Tung, and T.~Ebrahimi, ``Towards neural network approaches for
  point cloud compression,'' in \emph{Applications of Digital Image Processing
  XLIII}, vol. 11510.\hskip 1em plus 0.5em minus 0.4em\relax SPIE, 2020, pp.
  18--37.

\bibitem{9447226}
X.~Sheng, L.~Li, D.~Liu, Z.~Xiong, Z.~Li, and F.~Wu, ``Deep-pcac: An end-to-end
  deep lossy compression framework for point cloud attributes,'' \emph{IEEE
  Transactions on Multimedia}, pp. 1--1, 2021.

\bibitem{isik2021lvac}
B.~Isik, P.~A. Chou, S.~J. Hwang, N.~Johnston, and G.~Toderici, ``Lvac: Learned
  volumetric attribute compression for point clouds using coordinate based
  networks,'' \emph{arXiv preprint arXiv:2111.08988}, 2021.

\bibitem{wang2022sparse}
J.~Wang and Z.~Ma, ``Sparse tensor-based point cloud attribute compression,''
  \emph{arXiv preprint arXiv:2204.01023}, 2022.

\bibitem{huang2021hierarchical}
Y.~Huang, B.~Wang, C.-C.~J. Kuo, H.~Yuan, and J.~Peng, ``Hierarchical bit-wise
  differential coding (hbdc) of point cloud attributes,'' in \emph{ICASSP
  2021-2021 IEEE International Conference on Acoustics, Speech and Signal
  Processing (ICASSP)}.\hskip 1em plus 0.5em minus 0.4em\relax IEEE, 2021, pp.
  4215--4219.

\bibitem{yin2021lossless}
Q.~Yin, Q.~Ren, L.~Zhao, W.~Wang, and J.~Chen, ``Lossless point cloud attribute
  compression with normal-based intra prediction,'' in \emph{2021 IEEE
  International Symposium on Broadband Multimedia Systems and Broadcasting
  (BMSB)}.\hskip 1em plus 0.5em minus 0.4em\relax IEEE, 2021, pp. 1--5.

\bibitem{song2021layer}
F.~Song, Y.~Shao, W.~Gao, H.~Wang, and T.~Li, ``Layer-wise geometry aggregation
  framework for lossless lidar point cloud compression,'' \emph{IEEE
  Transactions on Circuits and Systems for Video Technology}, vol.~31, no.~12,
  pp. 4603--4616, 2021.

\bibitem{shannon1948mathematical}
C.~E. Shannon, ``A mathematical theory of communication,'' \emph{The Bell
  system technical journal}, vol.~27, no.~3, pp. 379--423, 1948.

\bibitem{8i}
``Common test conditions for {PCC},'' in \emph{{ISO}/{IEC} {JTC1}/{SC29}/{WG11}
  {MPEG} output document {N19324}}.

\bibitem{oord2016pixel}
\BIBentryALTinterwordspacing
A.~v.~d. Oord, N.~Kalchbrenner, and K.~Kavukcuoglu, ``Pixel {Recurrent}
  {Neural} {Networks},'' \emph{arXiv:1601.06759 [cs]}, Aug. 2016. [Online].
  Available: \url{http://arxiv.org/abs/1601.06759}
\BIBentrySTDinterwordspacing

\bibitem{salimans2017pixelcnn++}
\BIBentryALTinterwordspacing
T.~Salimans, A.~Karpathy, X.~Chen, and D.~P. Kingma, ``{PixelCNN}++:
  {Improving} the {PixelCNN} with {Discretized} {Logistic} {Mixture}
  {Likelihood} and {Other} {Modifications},'' \emph{arXiv:1701.05517 [cs,
  stat]}, Jan. 2017. [Online]. Available: \url{http://arxiv.org/abs/1701.05517}
\BIBentrySTDinterwordspacing

\bibitem{sullivan2012overview}
G.~J. Sullivan, J.-R. Ohm, W.-J. Han, and T.~Wiegand, ``Overview of the high
  efficiency video coding (hevc) standard,'' \emph{IEEE Transactions on
  circuits and systems for video technology}, vol.~22, no.~12, pp. 1649--1668,
  2012.

\bibitem{yuvtransform}
JVET, ``Ycgco-r: Observations and findings,'' in \emph{ITU-T SG 16 WP3 and
  ISO/IEC JTC 1/SC 29,20th Meeting, by teleconference, 7 – 16 Oct. 2020}.

\bibitem{loop2016microsoft}
C.~Loop, Q.~Cai, S.~O. Escolano, and P.~A. Chou,
  ``\BIBforeignlanguage{pt}{Microsoft voxelized upper bodies - a voxelized
  point cloud dataset},'' in \emph{\BIBforeignlanguage{pt}{{ISO}/{IEC}
  {JTC1}/{SC29} {Joint} {WG11}/{WG1} ({MPEG}/{JPEG}) input document
  m38673/{M72012}}}, May 2016.

\bibitem{d20178i}
E.~d'Eon, B.~Harrison, T.~Myers, and P.~A. Chou, ``\BIBforeignlanguage{ro}{8i
  {Voxelized} {Full} {Bodies} - {A} {Voxelized} {Point} {Cloud} {Dataset}},''
  in \emph{\BIBforeignlanguage{ro}{{ISO}/{IEC} {JTC1}/{SC29} {Joint}
  {WG11}/{WG1} ({MPEG}/{JPEG}) input document {WG11M40059}/{WG1M74006}}},
  Geneva, Jan. 2017.

\bibitem{usp}
\BIBentryALTinterwordspacing
M.~Zuffo, ``University of sao paulo point cloud dataset,'' (accessed Dec 19,
  2020). [Online]. Available: \url{http://uspaulopc.di.ubi.pt}
\BIBentrySTDinterwordspacing

\bibitem{mpegdataset}
M.~3DG, ``Preliminary dataset collection for ai-based point cloud
  experiments,'' in \emph{ISO/IEC JTC 1/SC 29/WG 7 N00326, Apr 2022.}

\bibitem{chang2015shapenet}
A.~X. Chang, T.~Funkhouser, L.~Guibas, P.~Hanrahan, Q.~Huang, Z.~Li,
  S.~Savarese, M.~Savva, S.~Song, H.~Su \emph{et~al.}, ``Shapenet: An
  information-rich 3d model repository,'' \emph{arXiv preprint
  arXiv:1512.03012}, 2015.

\bibitem{behley2019semantickitti}
J.~Behley, M.~Garbade, A.~Milioto, J.~Quenzel, S.~Behnke, C.~Stachniss, and
  J.~Gall, ``Semantickitti: A dataset for semantic scene understanding of lidar
  sequences,'' in \emph{Proceedings of the IEEE/CVF International Conference on
  Computer Vision}, 2019, pp. 9297--9307.

\bibitem{mentzer2018conditional}
\BIBentryALTinterwordspacing
F.~Mentzer, E.~Agustsson, M.~Tschannen, R.~Timofte, and L.~V. Gool,
  ``\BIBforeignlanguage{en}{Conditional {Probability} {Models} for {Deep}
  {Image} {Compression}},'' in \emph{\BIBforeignlanguage{en}{2018 {IEEE}/{CVF}
  {Conference} on {Computer} {Vision} and {Pattern} {Recognition}}}.\hskip 1em
  plus 0.5em minus 0.4em\relax Salt Lake City, UT: IEEE, Jun. 2018, pp.
  4394--4402. [Online]. Available:
  \url{https://ieeexplore.ieee.org/document/8578560/}
\BIBentrySTDinterwordspacing

\bibitem{theis2015generative}
L.~Theis and M.~Bethge, ``Generative image modeling using spatial lstms,'' in
  \emph{Advances in Neural Information Processing Systems}, 2015, pp.
  1927--1935.

\end{thebibliography}

\newpage

 \begin{wrapfigure}{l}{25mm} 
    \includegraphics[width=1in,height=1.25in,clip,keepaspectratio]{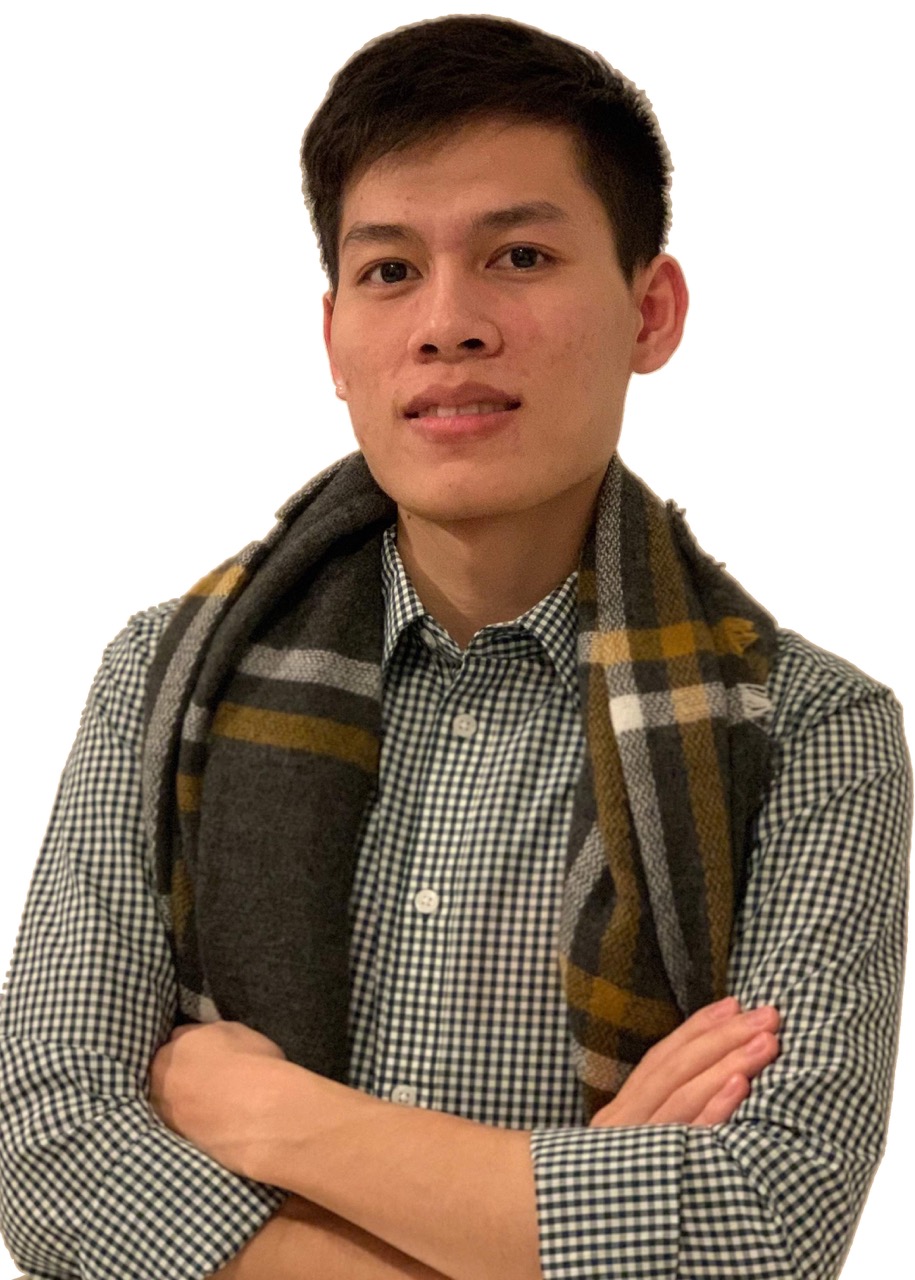}
  \end{wrapfigure}\par
  \textbf{Dat Thanh Nguyen}  received the Engineer degree in Electronics and Telecommunications from Hanoi University of Science and Technology, Vietnam and the M.S. degree in Electronics from the Polytechnic Institute of Paris, France. He has been working as an intern, first, and as research engineer, at L2S from May 2020 to May 2021. He is currently a PhD candidate at at Chair of Multimedia Communications and Signal Processing, Friedrich Alexander-Universität Erlangen-Nürnberg (FAU), Germany. From Jan 2023 to March 2023, he works as a PhD Research Scientist Intern at Meta Reality Labs, United States.
  
\vspace{0.6cm}
\setlength\intextsep{0pt}
\begin{wrapfigure}{l}{25mm} 
\includegraphics[width=1in,clip,keepaspectratio]{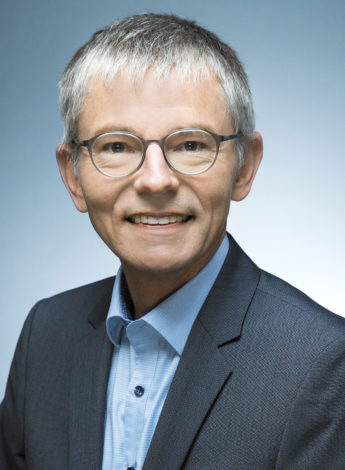}
\end{wrapfigure} \par
\textbf{André Kaup} (Fellow, IEEE) received the Dipl.-Ing. and Dr.-Ing. degrees in electrical engineering from RWTH Aachen University, Aachen, Germany, in 1989 and 1995, respectively.
\par He joined Siemens Corporate Technology, Munich, Germany, in 1995, and became the Head of the Mobile Applications and Services Group in 1999. Since 2001, he has been a Full Professor and the Head of the Chair of Multimedia Communications and Signal Processing, FriedrichAlexander-Universität Erlangen-Nürnberg (FAU),
Germany. From 2005 to 2007, he was a Vice Speaker of the DFG Collaborative Research Center 603. From 2015 to 2017, he served as the Head for the Department of Electrical Engineering and the Vice Dean for the Faculty of Engineering, FAU. He has authored around 400 journals and conference papers and has over 120 patents granted or pending. His research interests include image and video signal processing and coding, and multimedia communication.
\par Dr. Kaup is a member of the IEEE Image, Video, and Multidimensional Signal Processing Technical Committee a and the Scientific Advisory Board of the German VDE/ITG. In 2018, he was elected as a Full Member with the Bavarian Academy of Sciences. He was a Siemens Inventor of the Year 1998 and received the 1999 ITG Award and several IEEE Best Paper Awards. His group won the Grand Video Compression Challenge from the Picture Coding Symposium 2013. The Faculty of Engineering with FAU and the State of Bavaria honored him with Teaching Awards in 2015 and 2020, respectively.
He served as an Associate Editor for \textsc{ieee transactions on circuits and systems for video technology}. He was a Guest Editor for \textsc{ieee journal of selected topics in signal processing}.\par

\end{document}